\documentclass{aa} 
\usepackage{graphicx}
\usepackage{float}
\usepackage{tabularx}
\usepackage[utf8]{inputenc}
\usepackage{txfonts}
\usepackage[T1]{fontenc}
\usepackage{tgcursor}
\usepackage[colorlinks=true,linkcolor=blue,citecolor=blue]{hyperref}
\usepackage{appendix}
\usepackage{xcolor}
\DeclareRobustCommand{\ion}[2]{%
\relax\ifmmode
\ifx\testbx\f@series
{\mathbf{#1\,\mathsc{#2}}}\else
{\mathrm{#1\,\mathsc{#2}}}\fi
\else\textup{#1\,{\mdseries\textsc{#2}}}%
\fi}

\usepackage{natbib}
\setcitestyle{authoryear,open={((},close={))}}
\bibpunct{(}{)}{;}{a}{}{,} 

\begin{document}

   \title{Hubble constant and nuclear equation of state from kilonova spectro-photometric light curves}

   \author{M. A. P\'erez-Garc\'ia$^1$, L. Izzo$^2$, D. Barba-Gonz\'alez$^1$, M. Bulla$^3$, A. Sagu\'es-Carracedo$^4$, E. P\'erez$^5$, C. Albertus$^1$, S. Dhawan$^6$, 
   F. Prada$^5$, A. Agnello$^2$, C. R. Angus$^2$, S. H. Bruun$^2$, C. del Burgo$^7$, C. Dominguez-Tagle$^5$, C. Gall$^2$, \\A. Goobar$^4$, J.  Hjorth$^2$, D. Jones$^{8,9}$, A. R. López-Sánchez$^{10,11,12}$ and J. Sollerman$^3$}

   \institute{$^1$ Department of Fundamental Physics and IUFFyM, Universidad de Salamanca, Plaza de la Merced s/n 37008 Salamanca, Spain\\
   $^2$ DARK, Niels Bohr Institute, University of Copenhagen, Jagtvej 128, 2200 Copenhagen, Denmark\\
$^3$ The Oskar Klein Centre, Department of Astronomy, Stockholm University, AlbaNova, SE-10691 Stockholm, Sweden\\
$^4$ The Oskar Klein Centre, Department of Physics, Stockholm University, AlbaNova, SE-10691 Stockholm, Sweden\\
$^5$ Instituto de Astrof\'\i sica de Andaluc\'\i a (CSIC), Glorieta de la Astronom\'\i a, 18008 Granada, Spain\\
$^6$ Institute of Astronomy and Kavli Institute for Cosmology, University of Cambridge, Madingley Road, Cambridge CB3 0HA, UK\\
$^7$ Instituto Nacional de Astrof\'\i sica, \'Optica y Electr\'onica, Luis Enrique Erro 1, Sta. Ma. Tonantzintla, Puebla, CP 72840, Mexico\\
$^8$ Instituto de Astrof\'isica de Canarias, E-38205 La Laguna, Tenerife, Spain\\
$^{9}$ Departamento de Astrof\'isica, Universidad de La Laguna, E-38206 La Laguna, Tenerife, Spain\\
$^{10}$Australian Astronomical Optics, Macquarie University, 105 Delhi Rd, North Ryde, NSW 2113, Australia\\
$^{11}$Macquarie University Research Centre for Astronomy, Astrophysics \& Astrophotonics, Sydney, NSW 2109, Australia\\
$^{12}$Australian Research Council Centre of Excellence for All Sky Astrophysics in 3 Dimensions (ASTRO-3D), Australia\\
}

   \date{Accepted XXX. Received YYY; in original form ZZZ}

 
  \abstract
{The  merger of two compact objects of which at least one is a neutron star is signalled by
transient electromagnetic emission in a kilonova (KN). This event is accompanied by gravitational waves and possibly other radiation messengers such as neutrinos or cosmic rays. The electromagnetic emission arises from the radioactive decay of heavy $r-$process elements synthesized in the material ejected during and after the merger. In this paper we show that the analysis of KNe light curves can provide cosmological distance measurements and constrain the properties of the ejecta. In this respect, MAAT, the new Integral Field Unit in the OSIRIS spectrograph on the $10.4$ m Gran Telescopio CANARIAS (GTC), is well suited for the study of KNe by performing absolute spectro-photometry over the entire $3600-10000\hspace{2pt}\rm \AA$ spectral range. Here, we study the most representative cases regarding the scientific interest of KNe from  binary neutron stars, and we evaluate the observational prospects and performance of MAAT on the GTC to do the following: a) study the impact of the equation of state on the KN light curve, and determine to what extent bounds on neutron star (NS) radii or compactness deriving from KN peak magnitudes can be identified and b) measure the Hubble constant, $H_0$, with precision improved by up to 40$\%$, when both gravitational wave data and photometric-light curves are used. In this context we discuss how the equation of state, the viewing angle, and the distance affect the precision and estimated value of $H_0$.}
  

\titlerunning{$H_0$ and the nuclear EoS from KN spectro-photometric light curves.}
\authorrunning{M. A. P\'erez-Garc\'ia, L. Izzo, D. Barba-Gonz\'alez et al.}

\maketitle
%
\section{Introduction}
\label{intro}

The recent era of multimessenger astrophysics has opened up new possibilities to disentangle the properties of matter and probe the evolution of the Universe itself. One of the most prominent multimessenger observations was the {joint} detection of the gravitational wave (GW) event GW170817 on 2017 August 17.53 UT. Advanced LIGO/Virgo made the first detection of GWs from a binary neutron star (BNS)  merger \citep{abbott2017a}, with a simultaneous short gamma ray burst (GRB) detected by Fermi and INTEGRAL (GRB\,170817A, \citealt{goldstein2017,blackburn}). At $\sim$~0.5 days after the GW signal, an electromagnetic (EM) counterpart — the first for any GW source —  also known as a kilonova (KN), was identified within the $\sim$30 $\rm deg^2$ localization region and named AT 2017gfo. The source, located in the galaxy NGC 4993, was compatible with a BNS merger event at 40 Mpc \citep{distance,Hjorth2017}. The KN thermal emission observed in AT 2017gfo was produced by radioactive decay of neutron rich matter from the material ejected during and after the merger. 

In this paper we focus on compact star systems, binaries of neutron stars (NSs) that merge after the final evolutionary stages of inspiral coalescence. In these violent events, where gravity is in the strong regime, the merging process leads to the formation of a highly massive object, producing a broad multimessenger signal: GWs, EM waves, neutrinos \citep{neutrinosBNS} and even cosmic rays \citep{crBNS}. The dynamics of this kind of merger events is not yet well understood,  relying on computational simulations that include  detailed microphysics input at an increasing level of accuracy \citep{Dietrich17}. More importantly, neutrinos and weak interactions determine the composition and lepton fraction and ultimately the opacity of the expanding ejecta. This response modulates the nontrivial nonhomogeneous blue and red emission in the KN, as computational simulations have shown (see \citealt{metzger} and \citealt{Nakar2020} for recent reviews). The high-density material in BNS mergers is heated and  thermal neutrinos with energies in excess of a few MeV are expected \citep{neutrinoT}. Direct detection of thermal neutrinos by existing water-tank detectors such as Super-Kamiokande \citep{superK} is quite challenging for mergers at a distance $\gtrsim$100 Mpc; however, it is likely that they will contribute to the diffuse neutrino background. High-energy neutrinos associated with the GW170817 merger were also searched for by ANTARES, IceCube, and Pierre Auger Observatories, but no significant signal was observed \citep{albert}.

The KN coincident with GW170817 showed  a rapidly fading EM transient in the optical and infrared bands. The term kilonova was  historically coined from the perception of its brightness being a thousand times larger than a nova \citep{Li1998,Rosswog2005,Metzger2010}. We discuss in this work  that, even if far from such a standardized luminosity, clear dependencies on physical BNS quantities can be extracted. For AT 2017gfo, the observed spectral evolution is suggested to arise from a mixed composition of $r-$process elements in the ejected material \citep{Gillanders2022}; while early-time spectra are consistent with light $r-$process elements (nuclear masses  $A\sim$ 90–140), later spectra require intermediate composition, producing even heavier elements such as lanthanides. An alternative scenario, such as dust formation
in the KN that could also explain such a blue-red evolution, has been 
ruled out \citep{Gall17}. In addition, a broad feature observed at $\sim0.7-1\,\mu$m has been interpreted as due to \ion{Sr}{ii}
\citep{Watson2019,Domoto2021,Gillanders2022}. Since then, no other KN has been firmly detected (see e.g., \citealt{CoughlinNatAstro} and references therein for a summary of the follow-up efforts during LIGO and Virgo's third observing campaign). Another merger candidate is linked to GW190814, located farther away at $\sim 240$ Mpc, whose origin remains uncertain. However, it lacks of any EM counterpart and could likely be caused by a binary black hole merger \citep{essick, tews}. 
{GW190425 was the first BNS candidate in O3. The LALInference localization pipeline \citep{Veitch2015PhRvD..91d2003V} provided a 90\% credible region of 7461 deg$^2$, estimated luminosity distance of 156 $\pm$ 41 Mpc, much larger than that for GW170817, which complicated the search campaign resulting in no EM counterpart identified. The inferred merger parameters from the GW analysis published in \cite{abbott190425} for chirp mass $1.44^{+0.02}_{-0.02} \ \rm M_{\odot}$ are the following. Total mass $M_{1}+M_{2} \simeq 3.4_{-0.1}^{+0.3} M_{\odot}$, with $M_{1} \in(1.62,1.88) M_{\odot}, M_{2} \in(1.45,1.69) M_{\odot}$, and $\tilde{\Lambda} \leq 600$ for low spin prior, or $M_{1} \in(1.61,2.52) M_{\odot}, M_{2} \in(1.12,1.68) M_{\odot}$, and $\tilde{\Lambda} \leq 1100$ for high-spin prior.
Note that for the other confirmed BNS event,  GW170817, the inferred total mass is $M_{1}+M_{2} \simeq 2.73_{-0.01}^{+0.04} M_{\odot}$ with $M_{1} \in(1.36,1.60) M_{\odot}, M_{2} \in(1.16,1.36) M_{\odot}$, and $\tilde{\Lambda}=300_{-230}^{+420}$ for low-spin prior \citep{abbott2019}.}

Detecting KNe as EM counterparts to BNS mergers can shed light on different relevant areas of physics. For instance, it is expected that tighter constraints on existing cosmological models, formation of elements, the internal composition of the individual NSs, and even more fundamental physics can benefit from present and future multimessenger observations. 

One particular example is the constraints on the Equation of State (EoS) as it \mbox{describes} the interior of the NSs. KN observables such as spectra and light curves are controlled by the properties of the material ejected during the merger, including mass, expansion \mbox{velocity}, geometry and composition
(setting the opacity). These, in turn, depend on the properties of the merging system like e.g. NS and BH mass/spin and NS radius or the ratio of both, i.e. compactness. 
Constraints on NS masses and radii have been provided in a series of world-wide experimental efforts. In particular, recent observations from the Neutron star Interior Composition ExploreR (NICER) mission suggest that the EoS of NS might be rather stiff \citep{Miller2021, Raaijmakers2021, Riley2021}, indicating that many NS mergers should eject a significant amount of material and produce bright KNe.
From the population \mbox{studies} of Galactic double NSs (DNSs) it appears that $\simeq98\%$ of all merging DNSs will have a mass ratio, $q$, close to unity, $q<1.1$ \citep{zhang}. Recent studies find that the weighted mean masses of the primary and companion stars are respectively $(1.439\pm0.036) M_\odot$ and $(1.239\pm0.020) M_\odot$ \citep{yiyan}. In addition, the surface magnetic field strength in the primary stars of the DNSs is $B\sim 10^{10}$ G, and the spin period is $P\sim 50$ ms. Due to the fact that the NS in the BNS events under scrutiny will share these general characteristics it is important to see what information the EM counterparts can  yield.

One other example is linked to the value of the Hubble constant $H_0$, especially interesting in the context of the present tension in $H_0$ measurements from the Early \citep{Planck2020} and the Late \citep{Riess2021} Universe (see e.g. \citealt{Verde2019} for a recent review). The simultaneous detection of GWs and EM radiation from the same source led to independent measurements of the distance and redshift of the source, thus providing an estimate of $H_0$ that is independent of cosmological model assumptions as well as the local distance ladder \citep{distance}. Using GW as ``standard sirens'' \citep{Holz2005}, this
approach holds promise to arbitrate the existing tension in $H_0$ measurements. However,
the well-known degeneracy in the GW
signal between the \mbox{luminosity} distance and inclination
angle translates into large uncertainties on $H_0$, even when an EM counterpart is
identified \citep{distance}. Independent estimates of the viewing angle of GW170817 from either the GRB afterglow \citep[e.g.][]{Guidorzi2017}, superluminal motion of the GRB jet \citep{Hotokezaka2019} or the KN \citep{Dhawan2020,Coughlin2020,Dietrich2020} have been used to reduce the degeneracy between inclination and luminosity distance and therefore the uncertainties on $H_0$ ({see \citealt{Bulla2022} for a recent review} and e.g. \citealt{NakarPiran2021} and \citealt{Heinzel2021} for possible systematics introduced by these approaches).
Given the rapid evolution of KNe, and their intrinsic faintness, a prompt activation of ground-based telescopes is needed to get spectral information at their early epochs. 

In addition to currently operating spectrographs on 8-10 meter class telescopes, a new facility will be promptly available to investigate the nature of KNe. Here we present and discuss the role that MAAT\footnote{http://maat.iaa.es} (Mirror-slicer Array for Astronomical Transients) will have on KN science. MAAT is a new Integral Field Unit (IFU), based on image slicers, for the OSIRIS spectrograph on the 10.4-m telescope on the Gran Telescopio de CANARIAS (GTC, La Palma), which is planned to being operations in autumn 2023 \citep{maatwp}. Thanks to the pointing accuracy and the large collecting area of the GTC telescope, MAAT will permit to obtain a spectrum of a KN candidate in less than five minutes from its alert activation. The integral-field spectroscopy capabilities of MAAT, combined with its field-of-view (8.5$^{\prime \prime}$ $\times$ 12.0$^{\prime\prime}$) will allow to collect all the photons incoming from KN emission over the wavelength range $3600-10000$ $\AA$, including its immediate galaxy host environment, avoiding any possible flux losses due to the narrow slit aperture in long-slit spectrographs. Hence, MAAT is an unique capability in 8-10m class telescopes world-wide that will allow to perform absolute spectro-photometry of KNe, and any other interesting transients, see \citep{maatwp}.

In this work we perform a detailed analysis on how \mbox{different} scenarios involving BNS could help understand not only the physics governing these events, including the nuclear matter EoS, but also how fundamental parameters in cosmology such as the Hubble constant, $H_0$, can benefit from complementary EM data obtained with MAAT. For this task we use the output of radiative transfer simulations of the KN yield in the UV-IR range for selected cases of BNS mergers as we discuss below.

The structure of this paper is as follows. In Section~\ref{sec:obs}, we discuss the expected features of future KNe observations and existing rates of GW detections linked to BNS rates. We also comment on the observational strategy to optimally capture the light curve and spectral features providing key information on the physical parameters involved. 
In Section~\ref{sec:modelsandEOS}, we explain the subset of adopted EoSs in  detail and set up the scenario for the analysis we later perform based on a sample of 23 KN runs and their emission light curves obtained with \texttt{POSSIS}, a  radiative transfer Monte Carlo code. We also analyze the impact of the EoS on ejecta magnitudes in light of additional complementary probes from GW observations. In Section \ref{ref:KNlc}, we analyze and discuss the shape of KN light curves considering aspects such as inclination angles modulating the measured signal. In Section \ref{H0}, we tackle the $H_0$ determination and sources of uncertainty providing some paradigmatic examples of MAAT performance to this task. Finally, in Section \ref{conclusions}, we present a discussion and draw our conclusions.

\section{Observations and binary neutron star merger rates}
\label{sec:obs}

In the BNS scenario, the understanding of the impact of the ejecta on the KN brightness and its usability are crucial to reduce the uncertainties on the Hubble constant of EoS itself. To what extent these light curves can be standardized \citep{kashyap, Coughlin} and many of the aspects of the modelling of the emission \citep{metzger,Radice} is still a matter of debate. 

The KN emission is expected to be preceded by GWs detected during the subsequent observing runs of the LIGO/Virgo/KAGRA (LVK) collaboration. Second generation (2G) Advanced LIGO \citep{aasi2015a}, Advanced Virgo \citep{acernese2015} and KAGRA, and the proposed Einstein telescope (ET) interferometers \citep{Maggiore2020} will be crucial in providing the initial alert for newly-discovered BNS systems. For the incoming run O4, LIGO is expected to detect BNS mergers up to 160-190 Mpc, whereas Virgo sensitivity limits to 80-115 Mpc\footnote{The most updated forecasts by the LVK collaboration can be found at https://dcc-lho.ligo.org/LIGO-P1200087/public}. KAGRA will start observing with a very limited sensivitity of 1 Mpc, which is expected to improve to $\sim$ 10 Mpc towards the end of the run. For the following O5 science run, beyond 2028, LIGO projects a sensitivity goal of 325 Mpc while Virgo and KAGRA project a target sensitivity of 260 Mpc and 128 Mpc, respectively. In addition, ET will detect binary systems containing NSs up to redshifts corresponding to the peak of the cosmic star-formation rate. This will allow the study of the broad field of formation, evolution and physics of NSs in connection with KNe and short GRBs, along with the star formation history and the chemical evolution of the Universe.

The detection of a short GRB event, coincident within the error box localization regions provided by the above GW detectors, will further support the nature of the detection and constrain further those error boxes where candidates will be searched, similarly to what happened for the AT 2017gfo event \citep{goldstein2017}. { Recent advanced KN population synthesis models predict a rate of joint GW and short GRB detection from a BNS merger between $\sim$ 1-6 events per year \citep{Colombo2022,Patricelli2022}}. During the LIGO-Virgo (LVC)  O1, O2 and O3 runs, around $\sim$50  candidates GW events were reported, all regarding detections of compact objects binary mergers \citep{abbott2021prx}. One key aspect when considering the statistics available for any further study is the local merger rate density. The BNS rate is estimated to be $10-1700$ $\rm{Gpc}^{-3}\,\rm{yr}^{-1}$ as reported in the latest catalog from LVC \citep[GWTC-3,][]{abbott2022}. From the analysis of GW170817, a BNS rate of $110-3840$ $\rm{Gpc}^{-3}\,\rm{yr}^{-1}$ was derived \citep{abbott2021prx}. More specifically, the number of BNS detections in O4 is expected to be $10^{+52}_{-10}$ \citep{abbott2020a}. { This number depends on several assumptions on BNS physical properties and detector characteristics. Using simulations of BNS mergers based on the results obtained during the last observing O3 run, \citet{Petrov2022} used different constraints on the minimum network signal-to-noise ratio (S/N) threshold, S/N $>$ 8, for detection of a BNS coalescence, and on the number of interferometers, only one, that would provide a detection of a BNS merger. In \citet{abbott2020a} a network SNR>12 was used and a detection in at least two interferometers of the LIGO-Virgo collaboration was assumed. These different assumption led to a much better agreement with the results of the O3 run, and provide for the incoming O4 run an expected annual rate of BNS mergers of 34$^{+78}_{-25}$ events, while for the following O5 run an annual rate of 190$^{+410}_{-130}$ events. }. 

The asymmetry in the NS masses and chirp mass in the BNS will also make an impact on the reachable depth, for example for O4 the LIGO range for BNS  with equal mass ratio $q=1$ and individual NS masses $M_1=M_2=1.4M_\odot$ is $\sim 190$ Mpc while for BNS having $q \ne 1$ and $M_1 \times M_2=1.4M_\odot  \times	2.0M_\odot$	one has $\sim 220$	Mpc \citep{abbott2020a}. 

However, once the detection by GW interferometers marks the occurrence of a BNS merger, the identification of its EM counterpart is more challenging. In order to capture the IR-optical-UV counterpart of a candidate BNS  merger, an alert system exists. It consists in a communication node, the  Gravitational-wave Candidate Event Database, called GraceDB\footnote{https://gracedb.ligo.org/} which connects LIGO and Virgo analysis pipelines and sends an alert to astronomers when a promising GW signal has been detected and located in the sky within a given region. That initiates a search campaign with telescopes to point in the highlighted sky area, trying to capture the EM counterpart before it has faded away and ideally before it peaks. From previous events, such as GW170817, the community has gained valuable experience for those to come in the future. Note that this GW event was localized to a sky area of around 30 $\rm deg^2$ within $\sim$ 10 hours from the first alert. For the follow-up of a posterior event, GW190814, it was narrowed to about 2 hours \citep{meerLICHT} and  sky localization for this event was ameliorated to $18.5$ $\rm deg^2$ \citep{abbott2020b}. 
In the same line, for O3 and O4 the localizations of BNS area {were expected to be} $270^{+34}_{-20}$ and $33^{+5}_{-5}$ (in units of deg$^2$, $90\%$ c.r.) respectively, noting that for the later value the Hanford-Livingston-KAGRA-Virgo (HLKV) is included. In reality, the localizations during O3 were around 1000 to 10000 deg$^2$ due to low significant trigger alerts and single detector detections, and at distances $\sim$5 times farther than GW170817 \citep{kasliwal2020}. 

As current KN emission light curves obtained by simulations have shown, the brightness in the KN bolometric luminosity curve peaks at a time, $t_{\rm peak}$, ranging between tenths of a day and a few days from the merger, depending on intrinsic  physical parameters (ejected mass, expansion velocity, composition). The strategy to perform observations of a given KN candidate could be summarized as follows. As a first trigger alert will come from LVK with an estimate of distance and a sky localization area, an immediate search of a possible optical counterpart will begin. An estimate based on the two dimensional projection plus the distance uncertainty yields a three dimensional error box that could be set in order to localize the galaxies of interest for the EM follow-up (up to $\sim$300 Mpc) in existing catalogs such as the GLADE catalog \footnote{http://glade.elte.hu/}. Thereafter, a few candidates { observed within the LVK error region and spatially close to galaxies having distances within the range values reported by the BNS-GW merger signal might be detected by larger surveys such as the Zwicky Transient Facility (ZTF, \citealp{Bellm2019}), the Pan-STARRS \citep{Chambers}, the Asteroid Terrestrial-impact Last Alert System (ATLAS, \citealp{Tonry2018}), the Young supernova Experiment (YSE, \citealp{Jones2021}), or smaller size optical telescopes network such as the Gravitational wave Optical Transient Observatory (GOTO, \citealp{Dyer}), the Global Rapid Advanced Network Devoted to the multimessenger Addicts (GRANDMA, \citet{Antier}), the All Sky Automated Survey for SuperNovae (ASAS-SN, \citet{Shappee2014}), in addition to a multitude of other facilities, and then communicated to the astronomical community through Gamma-ray Coordinate Network (GCN) Notices and/or alert brokers, establishing the appearance of a potentially new KN candidate associated with the GW detection}. It is at this stage, when a recognition of the nature of the candidate (through its spectrum) is needed, that it is expected that MAAT will join the observation. 

\begin{figure}[t!]
    \centering
    \includegraphics[width=1.05\linewidth]{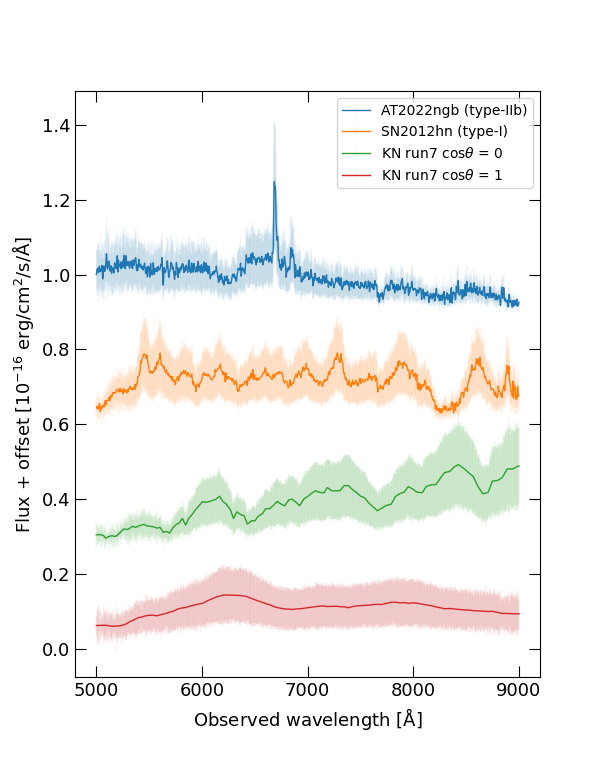}
    \caption{{Qualitative comparison of the spectra, and their errors, of (from the top): 1) the intrinsically faint type I SN 2012hn (at 4 days before maximum, \citet{Valenti2014}, note also the presence of bright nebular emission lines, such as H$\alpha$, [N II] and [S II], from the host galaxy); 2) the faint type IIb SN 2022ngb before peak luminosity \citep{Izzo2022}; 3) the modeled KN spectra of run7 viewed from edge on ($\cos \theta=0$), 4) and face on ($\cos \theta=1$) inclinations, as if they have been observed with MAAT/GTC using the grism R500 and an exposure time of 300s, and being characterized by an observed magnitude of $V=21.5$ mag.}
}
    \label{snkn}
\end{figure}

Additional transient sources such as intrinsically faint ($M_{peak} \sim -17 \to -15$ mag) supernovae Type I and II  will appear quite differently, see Fig. \ref{snkn}, where we compare spectra { as if they were observed by MAAT/GTC using the grism R500 and an exposure time of 300s (corresponding to the setup used for the KN search campaign, immediately following the GW detection alert)} of the fast declining type I SNe SN 2012hn (at 4 days before maximum \citealp{Valenti2014}) {and the faint type IIb SN 2019bkc (before peak luminosity \citealp{Izzo2022})} with the modeled KN spectra of run 7 (see Table \ref{tab1}) at face on and edge on inclinations. { All the spectra displayed in Fig. \ref{snkn} have been also simulated as if they had the same observee magnitude, namely $V = 21.5$ mag.} MAAT pointing will be characterized by a precision better than $\sim 1^{\prime \prime}$. 

\begin{figure}[ht!]
  \includegraphics[width=1.00\linewidth]{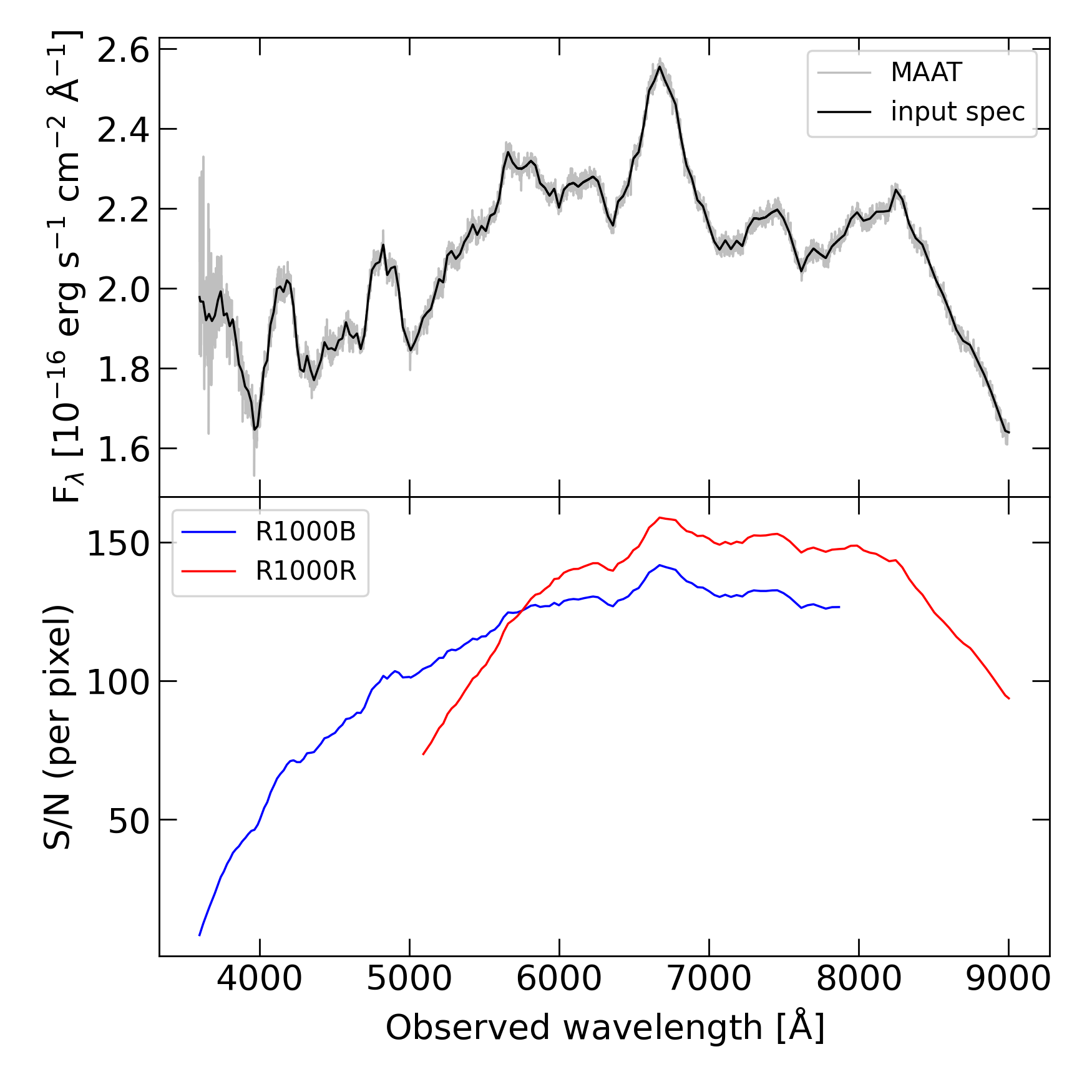}  \caption{Simulated KN spectral emission for an event similar to AT  2017gfo  at $\sim$ 1 day as observed with MAAT using the R1000B and R1000R gratings and a total exposure time of 1800s. The top panel shows the resulted observed spectra (grey) compared to the theoretical input from KN run 1 simulation (black) with EoS DD2, see Table \ref{tab1}, assuming similar parameters to those associated to GW170817 at 40 Mpc. The bottom panel shows the S/N obtained for the simulated event for gratings R1000B (blue) and R1000R (red). Note that the top spectrum is the combination of the flux captured by both red (R) and blue (B) R1000 gratings.}
  \label{fig1}
\end{figure}
\subsection{Emission patterns and photometry }

In order to achieve spectro-photometry with accuracy better than 3\% in the entire $3600-10000$ $\AA$ wavelength range available to MAAT, it will be necessary to observe spectro-photometric standard stars at different airmass.
Due to the slicer design of MAAT, there will be no slit losses due to seeing, atmospheric dispersion, or any other such effects that are a handicap to longslit observations. By performing multiepoch observations, additional features and their evolution can be observed that will facilitate the source/event identification.

\begin{figure}[ht!]
  \includegraphics[width=1.05\linewidth]{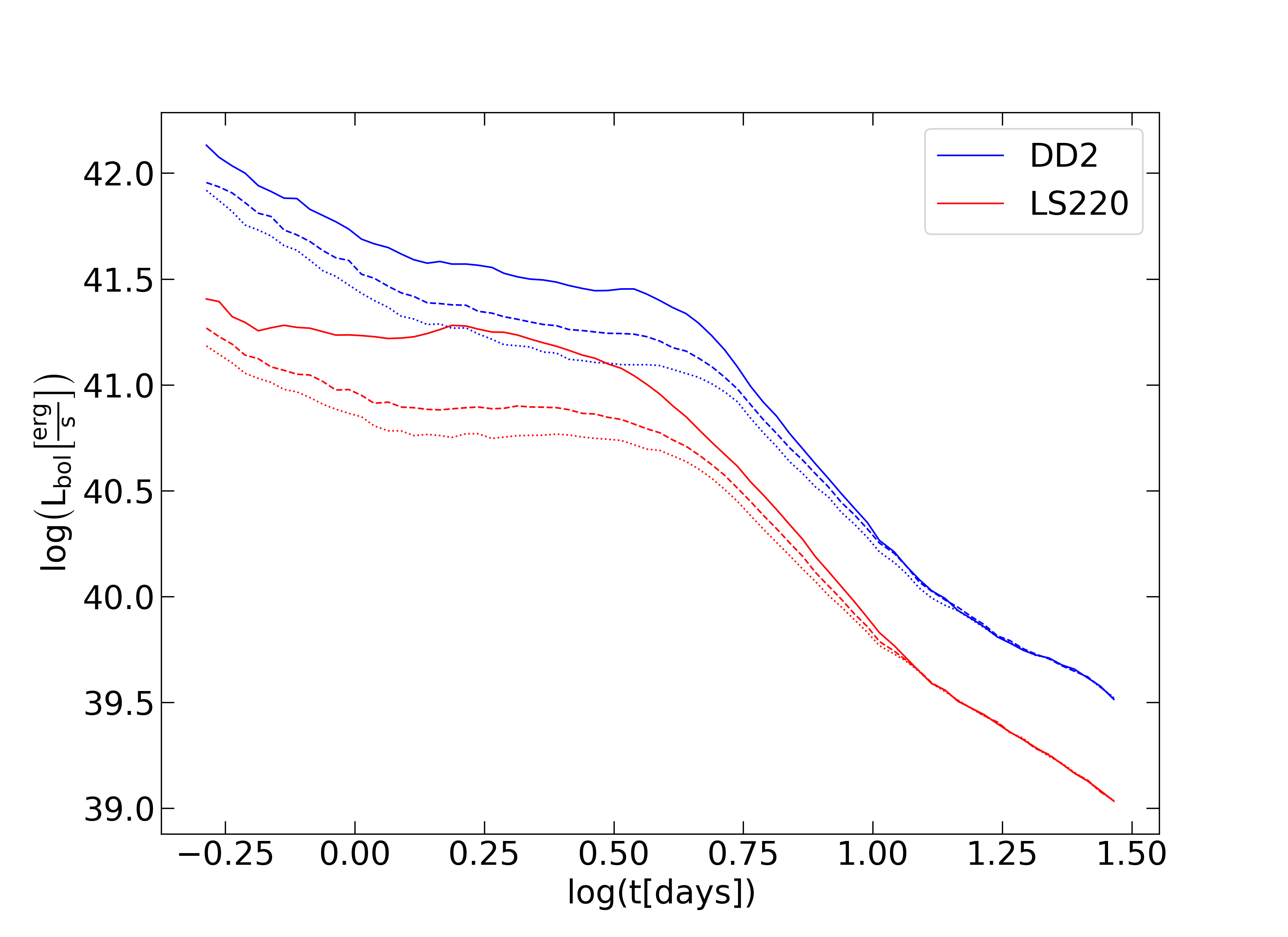}
   \caption{Luminosity as a function of time obtained from two selected simulations concerning run 1 with EoS DD2 (blue line) and run 2 with EoS LS220 (red line), see Table~\ref{tab1}. Solid, dashed and dotted lines refer to inclination angles $\cos\theta=1,0.7,0$, respectively.}
  \label{fig3}
\end{figure}

Figure \ref{fig3} shows the bolometric luminosity $L_{\rm bol}$ as a function of time obtained from two simulations concerning run 1 with EoS DD2 (blue line) and run 7 with EoS LS220 (red line), see Table~\ref{tab1} and Section \ref{sec:modelsandEOS} for more details.  Solid, dashed and dotted lines refer to polar angle inclinations $\theta$ corresponding to $\cos \theta=1,0.7,0$, respectively. As mentioned before, the observed light curves depend on intrinsic properties of the BNS and observing conditions. As for the former, the role of the EoS is of key importance, as it mostly determines matter interaction and governs the amount of mass that is ejected  during the tidal forces originated in the merger event. As shown in the figure, the inclination angle $\theta$ measured as a polar angle would allow to discriminate the EoS as emitted luminosities can differ  a factor of 10 for selected epochs. From  Fig. \ref{fig3} it is also clear that for a polar observer, luminosity will be larger as compared to any other direction. 
\begin{figure}[ht!]
  \includegraphics[width=1.0\linewidth]{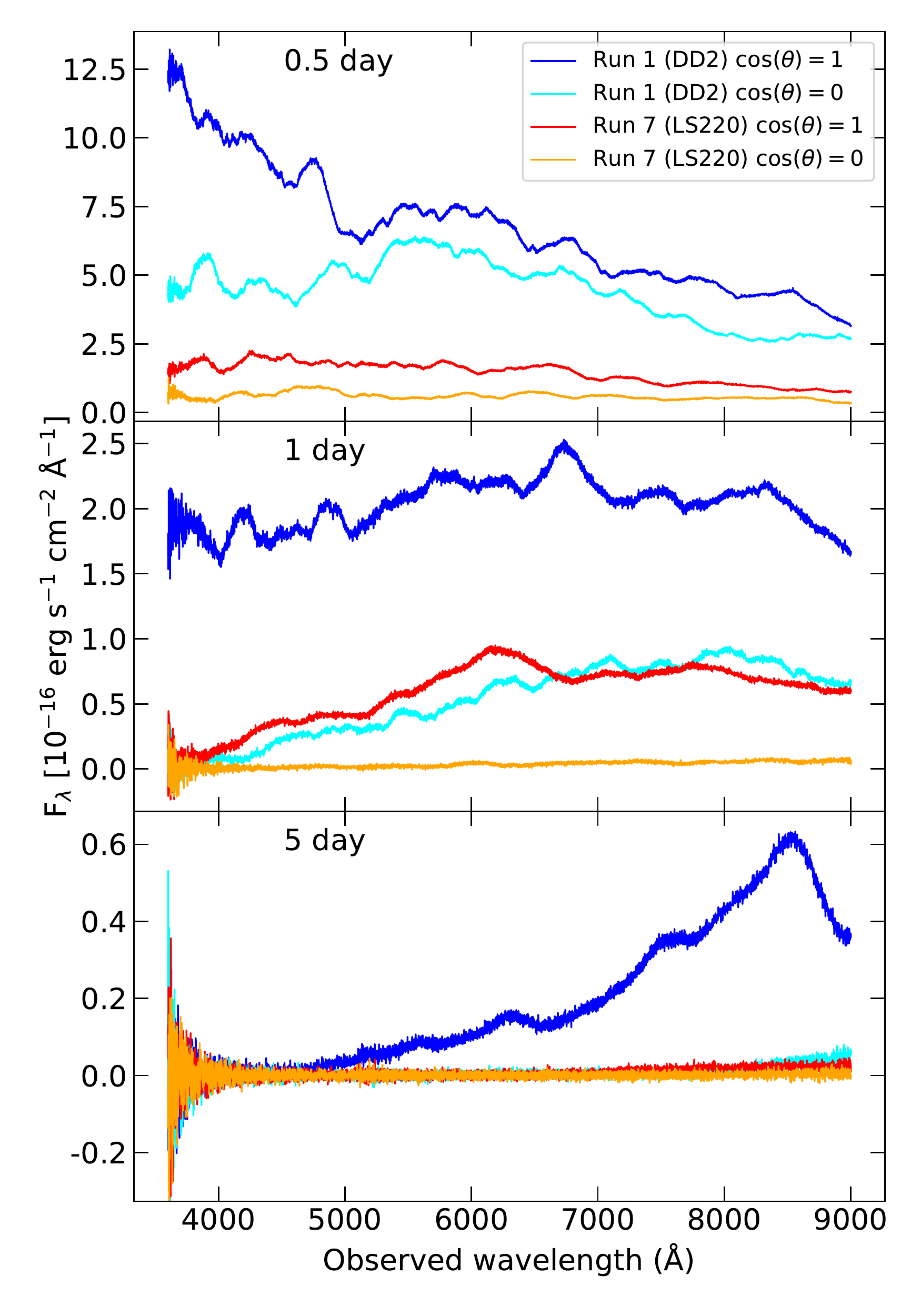}
     \caption{Absolute flux obtained  at $t=0.5,1,5$ days after the merger as seen by MAAT using R1000B and R1000R gratings. We plot run 1 (EoS DD2) for a polar and equatorial observer (dark and pale blue lines). Same for run 7 (EoS LS220) for a polar and equatorial observer (red and orange lines). A distance of 40 Mpc is assumed.}
 \label{fig4}
\end{figure}

In order to characterize the optical emission with MAAT, the  observational strategy will allow, provided optimal exposure times, to determine the spectral continuum, by fitting to a black body spectrum. Thus emission and absorption patterns due to the mixture of elements present in the ejected mass shells will show up. In the first stages of the KN emission, the dense environment will facilitate the appearance of blue-shifted absorption dips on top of the spectral continuum. Following the case of AT 2017gfo, for similar events, a \ion{Sr}{ii} triplet from neutron capture should appear toward the near-IR wavelength range \citep{Watson2019,Domoto2021,Gillanders2022}.
In this line, the analysis of the P-Cygni profiles will allow for the determination of the expanding velocity of the KN ejecta, as well as for abundance studies with spectral synthesis codes. It is important to remark that deviations from a simplified spherical geometry are likely to appear. In addition, a shock heated and disk wind structure form determining the blue KN  component, whose maximum radiation peaks in 1$-$2 days after the merger. In the same way, the red component will show up from days to weeks. Therefore a continuous and careful  monitoring is crucial to determine the peculiarities of the EM emission event and extract meaningful accurate data. Other features are also expected from different elements appearing in the neutron rich debris of the BNS \citep{Gillanders2022}. 

Figure \ref{fig4} shows the  absolute flux ${F}_\lambda$ at three epochs $0.5,1,5$ d (upper, middle and bottom panels) after the merger, as obtained by simulating the KN emission from a BNS merger at 40 Mpc with mass ratio $q=M_1/M_2=1$ and chirp mass $M_{\rm chirp}=\mathcal{M}=\left[\frac{q}{(1+q)^{2}}\right]^{3 / 5} M=1.22\,M_\odot$ where $M=M_1+M_2$. For this particular case $M_1=M_2=1.4\,M_\odot$. They correspond to run 1 with EoS DD2 with polar and equatorial observing angles (dark and pale blue lines) and run 7 with EoS LS220 (red and orange lines), see Table \ref{tab1} and Section \ref{sec:modelsandEOS}. The exposure time with MAAT has been set to $t_{\rm exp}=1800$ s. We can see that the effect of the EoS is most clearly seen at early times. We detail the role of the selected runs and EoS in Section \ref{sec:modelsandEOS}. The  viewing angle dependence must be carefully addressed, since it largely impacts the measured spectra (see below).

\subsection{Observation plans} 

We now detail how in the first phase of the identification of the KN emission, a series of low-resolution (R500B and/or R500R grisms, $R$ = 860-940) exposures will be used. With this setup, we can maximize both the telescope time and the number of candidates to check. This instrumental configuration will allow us to get spectra with a sufficient S/N (S/N $\gtrsim$ 10) to classify possible KN candidates up to $r \sim 21.5$ mag with a single exposure of 300s. The total amount of telescope time will be $\sim$ 1 hour to classify about ten candidates. Once, and if, a KN is identified {by our or other programs}, a monitoring program will be started with a daily cadence, using higher resolution grisms (e.g. R1000B/R, $R$ = 1630-1800) to trace the spectral continuum and identify spectral features attributable to newly-synthesised $r-$process elements \citep{Smartt2017,Watson2019}. {In Fig.~\ref{fig1}, we show the results of our simulations for a KN event similar to AT 2017gfo observed with MAAT using the R1000 grisms, $\sim$ 1 day after the GW signal detection.}
 
Although for some of the BNS events, an EM counterpart is expected be correlated with the merger event, as for AT 2017gfo and GW170817, for other events such as {GW190425} a counterpart was not detected, increasing the uncertainty about its dynamics and nature \citep{abbott190425}. 
The emission of associated GRBs could be observed with current orbiting detectors on-board satellites such as Fermi, Swift, INTEGRAL and the InterPlanetary Network (IPN),  a set of spacecrafts with gamma-ray detectors. This enables us to find out where on the sky an associated short GRB occurred using triangulation techniques \citep{Hurley2013}.  It may also happen that, to our regret, we may not observe any associated GRB, as only those beamed towards Earth are detectable. Placing the constraint that a jet is only launched if the mass of the remnant, $M_{\text {rem }}$, relates to the static TOV solution, $M_{\text {TOV }}$, as $M_{\text {rem }} \gtrsim 1.2 \times M_{\text {TOV }}$, leads different suites of combined models to predict a BNS jet-launching fraction of $f_{\mathrm{s}, \mathrm{BNS}} <0.6$ \citep{beamed}.

\section{Simulating KNe} 
\label{sec:modelsandEOS}

During a BNS merger, two NSs collide to finally produce thermal EM emission that we measure as a KN. Originally each NS is born as a hot lepton rich dense object evolving into a cold neutron rich one. Spinning magnetized stars are indeed possible and have been simulated \citep{boc}, although we work under the premise of low-spin zero magnetic field for them.  
The EoS is a key ingredient when describing the behavior of the matter inside NSs. The internal radial structure in the spherical static (non rotating) case is obtained after solving the Tolman-Oppenheimer-Volkoff (TOV) equations  providing the mass-radius relationship $M(R)$. Complementary measurements indicate NSs have a radius $R\lesssim14$ km and $M\lesssim 2.2\,M_\odot$, being the recently detected MSP J0740 $+6620$ the most massive NS to date  with a value $2.14_{-0.09}^{+0.10} M_{\odot}$ \citep{Cromartie19}. 
\begin{table*}[t!]
\centering
\begin{tabular}{|c|c|c|c|c|c|c|c|c|}
\hline
\multicolumn{1}{|l|}{Run} & EoS  & {\it q} & $M_{\rm chirp}\left(M_{\odot}\right)$ & $M_1\left(M_{\odot}\right)$ & $M_{\rm  wind}\left(M_{\odot}\right)$  & $M_{\rm dyn}\left(M_{\odot}\right)$ & $ \langle v_{\rm ej} \rangle/c$ & $\langle Y_{\rm e} \rangle $       \\ \hline
1                      & DD2  & 1 & 1.22         & 1.4 & $3.708 \times 10^{-2}$  & $4 \times 10^{-4}$  & 0.22 & 0.17   \\\hline
2                      & SFHo & 1 & 1.22  & 1.4 & $3\times 10^{-5}$  & $4\times 10^{-4}$        & 0.35 & 0.19  \\\hline
3                      & DD2  & 1 & 1.39         & 1.6 & $5.88\times 10^{-3}$  & $1.2\times 10^{-3}$  & 0.24  & 0.14    \\\hline
4                      & DD2  & 1 & 1.31         & 1.5 & $5.31\times 10^{-2}$  & $7.0\times 10^{-4}$  & 0.17 & 0.2  \\\hline
5                      & DD2  & 1.036 & 1.23        & 1.44 & $4.32\times 10^{-2}$  & $5.0\times 10^{-4}$  & 0.2 & 0.17  \\\hline
6                      & SFHo  & 1.036 & 1.23         & 1.44 & $2.7\times 10^{-4}$  & $4.0\times 10^{-4}$  & 0.33 & 0.18    \\\hline
7                      & LS220  & 1 & 1.22         & 1.4 & $1.37\times 10^{-2}$  & $1.4\times 10^{-3}$  & 0.17  & 0.14    \\\hline
8                      & LS220  & 1.036 & 1.23         & 1.44 & $1.17\times 10^{-2}$  & $1.9\times 10^{-3}$  & 0.16  & 0.14    \\\hline
9                      & LS220  & 1 & 1.31        & 1.5 & $4.80\times 10^{-4}$  & $3.0\times 10^{-4}$  & 0.19  & 0.08    \\\hline
10                      & LS220  & 1 & 1.39        & 1.6 & $2.10\times 10^{-4}$  & $3.0\times 10^{-4}$  & 0.21  & 0.07    \\\hline
11                      & LS220  & 1 & 1.49        & 1.71 & $1.80\times 10^{-4}$  & $3.0\times 10^{-4}$  & 0.22  & 0.08    \\\hline
12                        & SFHo & 1.092 & 1.137       & 1.365 & $2.64\times 10^{-2}$ & $1.5\times 10^{-3}$  & 0.23 & 0.14  \\\hline
13                       & SFHo & 1.17 & 1.128       & 1.4 & $3.52\times 10^{-2}$ & $1.2\times 10^{-3}$  & 0.2 & 0.14   \\\hline
14                       & SFHo & 1 & 1.175       & 1.35 & $1.87\times 10^{-2}$ & $3.50\times 10^{-3}$  & 0.24 & 0.17  \\\hline
A                        & LS220 & 1 & 1.188       & 1.365 & $2.16\times 10^{-2}$ & $1.6\times 10^{-3}$  & 0.16 & 0.22  \\\hline
B                        & DD2 & 1 & 1.188       & 1.365 & $4.62\times 10^{-2}$ & $1.1\times 10^{-3}$  & 0.18 & 0.25  \\\hline
C                        & SFHo & 1 & 1.188       & 1.365 & $5.67\times 10^{-3}$ & $2.8\times 10^{-3}$  & 0.21 & 0.23  \\\hline
D                        & DD2 & 1.43 & 1.188       & 1.637 & $9.12\times 10^{-2}$  & $7.0\times 10^{-3}$ & 0.14 & 0.14  \\\hline
E                        & LS220 & 1.43 & 1.188       & 1.637 & $5.34\times 10^{-2}$  & $7.3\times 10^{-3}$ & 0.17 & 0.16  \\\hline
F                        & LS220 & 1.66 & 1.188       & 1.769 & $2.04\times 10^{-2}$  & $1.11\times 10^{-2}$ & 0.14 & 0.07  \\\hline
G                        & DD2 & 1.22 & 1.188       & 1.509 & $6.27\times 10^{-2}$ & $2.50\times 10^{-3}$  & 0.17 & 0.19  \\\hline
H                        & SFHo & 1.43 & 1.188       & 1.637 & $6.03\times 10^{-2}$  & $3.80\times 10^{-3}$ & 0.2 & 0.14  \\\hline
I                        & SFHo & 1.66 & 1.188       & 1.769 & $5.31\times 10^{-2}$ & $1.50\times 10^{-3}$  & 0.12 & 0.07  \\\hline
\end{tabular}
\caption{Runs simulated and discussed in this paper. Column \textit{Run} is the label of the run, \textit{EoS} is the Equation of State assumed in the Numerical Relativity (NR) simulation, \textit{q}, $M_{\rm chirp}$ and $M_1$ are the mass ratio, chirp mass and mass of the heavier NS, respectively; $M_{\rm wind}$, $M_{\rm dyn}$  are the wind and dynamically ejected mass respectively (we take $M_{\rm wind}=0.3\,M_{\rm disk}$ as a prescription). $\langle v_{\rm ej} \rangle\equiv \langle v_{\rm dyn} \rangle$ and $\langle Y_{\rm e} \rangle \equiv \langle Y_{\rm e,dyn} \rangle $ are the average ejecta velocity and electron fraction composition in the dynamically ejected material. Models 1-14 correspond to \citealt{Radice} while A-I to \citealt{nedora}.}
\label{tab1}
\end{table*}


To study the impact on the $H_0$ values and description of nuclear matter EoS set by the KN physics, we have selected a set of 23 simulations in the literature (see Table~\ref{tab1}) for a reduced number of EoS. In particular, we select three different EoS: DD2 \citep{dd2}, LS220 \citep{ls220} and SFHo \citep{sfho}. We consider the works of \citet{Radice} and \citet{nedora}, {with models in the latter tailored to GW170817}. Note that the simulation of the outcome of these transients relies on all fundamental interactions known, i.e. gravitational (infall and large tides with object deformation), electroweak physics  (through the neutrino transport and nucleosynthesis yields) and hadronic (nuclear matter content of the NSs as described by the underlying EoS and fragmentation of ejecta and decay).  {Therefore, we choose these simulations as they include a} full neutrino transport and multidimensionality in order to see how the different ion population peaks and fractions appear in the spectra, for a discussion see \cite{tanaka2016}. 
It is important to note that some of these works miss a detailed treatment of neutrino emission/absorption, which may impact the ejecta masses in a systematic way, e.g. underestimating their value, and thus the KN light curve \citep{Radice}.
{The set of EoS used in our work is able to produce NSs with masses below the maximum static or TOV mass for the GWTC-3 analysis \citep{Abbott_2020c}. In this same study, when focusing on BNS binaries a broad, relatively flat NS mass distribution
extending from 1.2 to 2.0$M_\odot$ is obtained. Thus in this sense our subset of events can be safely considered to belong to this distribution. 

The situation regarding our choice of EoS is informed in this same sense by recent results obtained from Bayesian model selection on a wide variety of models using multimessenger observations. In Fig. 2 in \cite{Abbott_2020c} the Bayes factors for the narrow prior results using different waveform
approximants for GW170817 found that the EoS yielding excessively large tidal deformabilities are disfavored by the data. Therefore they are disfavored with a Bayes factor
that is much smaller when compared to the most favored EoS models. However, these authors find that only a few of the EoS used in the analysis can be ruled out as most of them have comparable evidences within an order of magnitude. They find this result by sorting the EoS by the compactness, C, of the neutron star at a fiducial mass of 1.36$M_\odot$. Therefore EoS producing more compact stars than those predicted for H4 with $C = 0.148$ are allowed, as it is the case for DD2, SFHo, LS220, with C = 0.152, 0.168 and 0.158, respectively. More in this line, \cite{Biswas_2022} used mass and tidal deformability measurements from two BNS  events, GW170817 and GW190425, and simultaneous mass–radius measurements of PSR J0030+0451 and PSR J0740+6620 by the NICER collaboration, being able to to rule out different variants of the MS1 family, SKI5, H4, and WFF1 EoS decisively among a set of 31 EoS. These EoS were found to be either extremely stiff or soft.    }

In Table~\ref{tab1}, we list some parameters that are key to understand the KN emission light curves as obtained from BNS merger simulations. We take $q$ and $\mathcal{M}$ as the mass ratio and chirp mass, respectively, $q=M_1/M_2$, $q\geq 1$, with individual NS masses $M_1, M_2$ and chirp mass $M_{\rm chirp}\equiv \mathcal{M}=\frac{\left(M_{1} M_{2}\right)^{3 / 5}}{\left(M_{1}+M_{2}\right)^{1 / 5}}$.
Extrinsic BNS parameters such as observing angle $\theta$, $q$ and  $M_{\rm chirp}$ are accessible through GW observation as well as the redshift $z$ and luminosity distance $D_{\rm L}$.
Intrinsic parameters are related to properties of the different contributions to the ejected material, in particular the mass from a dynamical ejecta component $M_{\rm dyn}$ and that from a wind component $M_{\rm wind}$ ejected from an accretion disk of mass $M_{\rm disk}$ created around the merger remnant. The ejecta are described by the average velocity, $\langle v_{ej} \rangle$, and the average electron fraction $\langle Y_{\rm e} \rangle$. In this scenario, matter is neutron rich (as neutrinos further deleptonize matter) so that $\langle Y_{\rm e} \rangle<0.5$.

\begin{figure*}[t]
\begin{center}
  \includegraphics[width=1.0\linewidth]{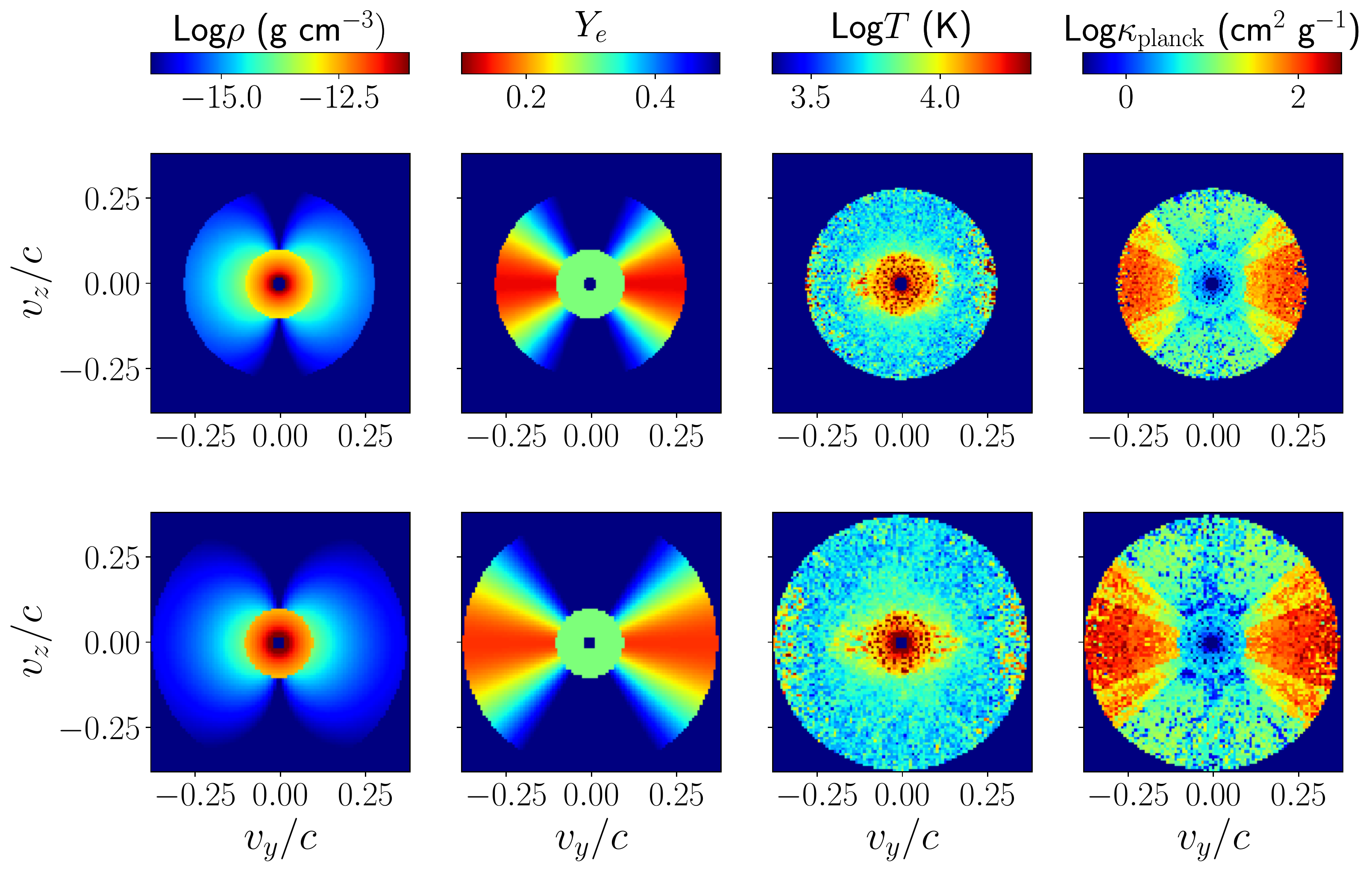}
   \caption{From left to right, distributions of mass density $\rho$, electron fraction $Y_{\rm e}$, temperature $T$ and Planck mean opacities $\kappa_{\rm planck}$ in the $v_y-v_z$ velocity plane. The top row refers to run A (LS220 EoS) and the bottom row to run B (DD2 EoS), both with $q=1$ and ${M}_{\rm chirp} = 1.188 \,M_{\odot}$, see Table \ref{tab1}. Maps are computed at an epoch of 1~d after the merger. The pixelization seen in the $T$ (and hence also $\kappa_{\rm planck}$) map is due to the temperature being computed with estimators in the code and thus being subject to Monte Carlo noise (see text for more details). Analytic functions are instead used for $\rho$ and $Y_{\rm e}$ and the distributions are therefore smoother.}
 \label{fig_rhoTrunAB}
 \end{center}
\end{figure*}

For each of the models in Table~\ref{tab1}, KN spectra are extracted using the 3D Monte Carlo radiative transfer code \texttt{POSSIS} \citep{Bulla2019}. The input geometries are characterized by a spherical disk-wind ejecta component at relatively low velocities surrounded by a dynamical ejecta component concentrated around the merger plane. Regarding the disk-wind component, we assume a spherical outflow extending from $v^\mathrm{wind}_\mathrm{min}=0.02$\,c to $v^\mathrm{wind}_\mathrm{max}=0.1$\,c (corresponding to an average velocity of 0.05\,c). {The composition and fraction of the disk that is ejected are highly uncertain \citep[e.g.][]{Just2015,Siegel2018,Fernandez2019,Miller2019b, Fujibayashi2020}, with numbers for the latter varying from roughly 20 to 40\%. For each model, here we fix the electron fraction to $Y_{\rm e,wind}=0.3$ and the wind mass to an average value of 30\% of the disk mass $M_\mathrm{disk}$, although future studies should explore the impact of these assumptions on the KN observables}. The density profile in the disk-wind component is assumed to scale with radius $r$ as $\propto {r}^{-3}$. For the dynamical ejecta component, we assume an outflow extending from $v^\mathrm{dyn}_\mathrm{min}=0.1$\,c to $v^\mathrm{dyn}_\mathrm{max}$ such that the mass-averaged velocity is the one required for each model, $\langle v_{\rm dyn} \rangle$. We set the mass to the specific value for each model and assume the $Y_{\rm e,dyn}$ distribution to have an angular dependence of the form $Y_{\rm e}\propto {\rm A}+{\rm B}\,\cos^2\theta$ (consistent with simulations in \citealt{Radice}, C. Setzer, private communication) where A and B are scaling constants chosen to match the desired $\langle Y_e \rangle \equiv \langle Y_{\rm e, dyn}\rangle$ in each model. A broken power law is chosen for the density profile in the dynamical ejecta, with density scaling with radial distance as $\propto r^{-4}$ below and as $\propto r^{-8}$ above 0.4\,c. The density profile in the dynamical ejecta component is modeled with an angular dependence $\propto\sin^2\theta$ that is found in numerical-relativity simulations \citep{Perego2017,Radice} and has been shown to provide a better fit to AT 2017gfo \citep{Almualla2021}. Figure~\ref{fig_rhoTrunAB} shows the distributions of mass density $\rho$, electron fraction $Y_{\rm e}$, temperature $T$ and Planck mean opacities $\kappa_{\rm planck}$ at 1~d after the merger for run A (LS220 EoS) and run B (DD2 EoS).

For each model, we run Monte Carlo radiative transfer simulations using \texttt{POSSIS} and extract spectra for $11$ viewing angles equally spaced in cosine from pole (face-on) to equator (edge-on) and for 100 epochs going from 0.1 to 30 days after the merger (with logarithmic binning of $\Delta\log t=0.0576$). In particular, we use a new version of \texttt{POSSIS} with state-of-the-art heating rate libraries \citep{submitted}, density-dependent thermalization coefficients \citep{Wollaeger2018} and wavelength- and time-dependent opacities from \cite{tanaka2020}. Assuming Local Thermodynamic Equilibrium (LTE), the temperature is initialized according to the local heating rates at the start of the simulation and then updated at each time-step via Monte Carlo estimators of the mean intensity in each cell \citep{Lucy2003}. In all the simulations, $10^6$ Monte Carlo photons are created and followed in their propagation throughout the grid as they interact with matter (electron scattering and bound-bound transitions are considered) and eventually leave the ejecta.
\begin{figure}[h!]
  \includegraphics[width=1.05\linewidth]{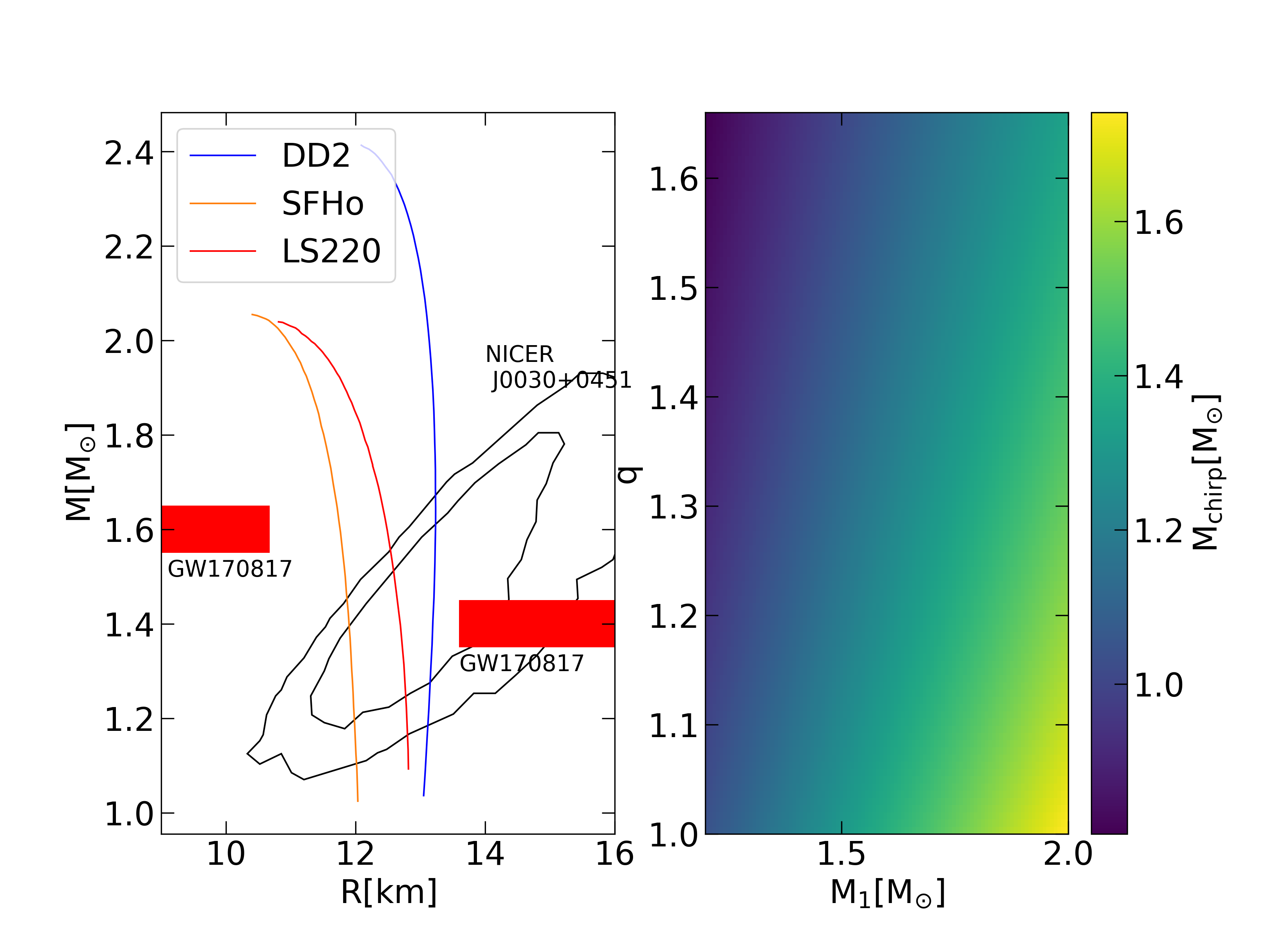}
   \caption{(Left panel) Mass-radius relation for the three EoS that we use in this work DD2 \citep{dd2}, SFHo \citep{sfho} and LS220 \citep{ls220}. We also include radii constraints from GW170817 (\citealt{Bauswein2017}; \citealt{Annala2018}) and NICER \citep{Miller2019}. (Right panel) Values of chirp mass as a function of mass ratio $q$ and the largest of the NS masses, $M_1$.}
  \label{Fig:q_mc_m1}
\end{figure}
\subsection{KN ejecta and EoS}
\label{KNEoS}

In order to study further dependencies of the KN light curves on the actual matter  properties and EoS, we now focus on the ejecta. Susbsequently from NR simulations of BNS mergers, the asymmetry in the mass ratio, $q$, total mass $M_1+M_2$ and the chirp mass $M_{\rm chirp}$ have a profound impact on the amount, composition  and velocity of the ejecta and this determines the luminosity of the KN. 
It is believed that due to the high compactness $C\sim GM/Rc^2$ of individual NSs and the deep gravitational potential well, matter ejected in the merger is essentially stripped off external layers \citep{bauswein}. 
Going further into the NS description, its layered structure can be mainly explained from two regions: an external crust and a central core. In the usual picture,  matter is well beyond nuclear saturation density in the homogeneous nucleon core, $\rho_0\sim 2.4\times 10^{14}$ $\rm g\,cm^{-3}$, while densities in the crust are typically less than $\sim 0.1\rho_0$. The crust 
is the most external, least gravitationally bound layer and is mostly constituted of so-called pasta phases \citep{pasta} in the inner crust and, more externally, of ionic lattices at lower densities in a degenerate lepton sea \citep{chamel}. The nature of the many-body approach to describe such relativistic quantum dense matter in NSs allows the large spread of EoS available in the literature, for reference see \citep{oertel}. As mentioned, most of the  NS mass is in the core in the form of  baryons, although if densities are in excess of $\sim 6 \times 10^{14}$ $\rm g cm^{-3}$ a quark star or a hybrid NS with a quark deconfined inner core is hypothesized \citep{witten, olek1}. The gravitational mass, $M$, and baryonic mass, $M^*$, are quantities that determine the binding energy of the NS and some universal relations between them are available in the literature \citep{Gao}, based on the radius of a $1.4 \,M_\odot$ NS, $R_{1.4}$, 
$M^{*}=M+R_{1.4}^{-1} \times M^{2}$ showing that for different EoS this parameter is genuinely different \citep{Dietrich17}. 

 From GW170817 some constraints appear for the masses, i.e. total mass $M \approx 2.74\, M_{\odot}$, and individual masses of $M_{1} \approx(1.36-1.6)\, M_{\odot}$ and $M_{2} \approx(1.17-1.36)\, M_{\odot}$. 
From this analysis marginalized over the selection methods, an upper bound was obtained on the effective tidal deformability assuming a low-spin prior, which
is consistent with the observed NS population. Let us remind that the binary tidal polarizability is defined as
\begin{equation}
\tilde{\Lambda} \equiv \frac{16}{13}\left[\frac{\Lambda_{1} M_{1}^{4}\left(M_{1}+12 M_{2}\right)+\Lambda_{2} M_{2}^{4}\left(M_{2}+12 M_{1}\right)}{\left(M_{1}+M_{2}\right)^{5}}\right].
\end{equation}
 The individual NS tidal deformabilities $(i=1,2)$ are $\Lambda_i=\frac{2}{3} k_{2,i} C_{i}^{-5}$ with $k_{2,i}$ the Love number \citep{flanagan, universelattimer}. Phenomenologically it has been found that for the EoS we consider $\Lambda_i=A C_{i}^{-6}$ with $A_{\mathrm{DD2}}=0.0112 \pm 0.0002$, 
$A_{\mathrm{SFHo}}=0.0099 \pm 0.0002$, $A_{\mathrm{LS220}}=0.0111\pm 0.0002$.

From GW170817 upper bounds were found $\Lambda \leq 800$ at the $90 \%$ confidence level \citep{abbott2017a}, which disfavors EoS that predict the largest radii stars. Tighter constraints were found in a follow up reanalysis \citep{abbott2019} with $\tilde{\Lambda}=300_{-230}^{+420}$ (using the $90 \%$ highest posterior density interval), under minimal assumptions about the nature of the compact objects. 
The fact that NSs are not point-like objects imply there are strong tides accelerating the inspiral (with initial frequency $f_{\mathrm{GW}}$ and producing a phase shift \citep{flanagan} in the form $\delta \Phi_{t}\simeq -\frac{117}{256} \frac{(1+q)^{4}}{q^{2}}\left(\frac{\pi f_{\mathrm{GW}} G \mathcal{M}_{chirp}}{c^{3}}\right)^{5 / 3} \tilde{\Lambda}$.  In this respect, radii were  further constrained for both NSs in GW170817 to $R_{1}=11.9_{-1.4}^{+1.4} \mathrm{~km}$ and $R_{2}=11.9_{-1.4}^{+1.4} \mathrm{~km}$ at the $90 \%$ credible level \citep{eosgw}.

To date there are dozens of EoS provided on phenomenological grounds typically using effective field theories invoking nuclear degrees of freedom \citep{oertel}. In this work we have selected a reduced set of three different EoS: DD2 \citep{dd2}, LS220 \citep{ls220} and SFHo \citep{sfho} that are compatible with maximum mass, being larger than $\sim 2\,M_\odot$, and the binary tidal polarizabilities deduced in GW170817 \citep{abbott2017a}. 

As a typical example of stellar configurations populating the binaries, Fig. \ref{Fig:q_mc_m1} shows on the left panel the TOV  solution to the mass-radius relation for the three EoS we use DD2, SFHo and LS220, along with excluded regions arising from GW170817 due to \citealt{Bauswein2017} (left) and \citealt{Annala2018} (right), and confidence regions from pulsar measurements from NICER \citep{Miller2019}. { Since the BNS cases we analyze correspond to specific values of individual NS parameters computed with our set of EoS we show in the right panel the values of chirp mass as a function of mass ratio $q>1$ and the largest of the NS masses, $M_1$}.

We now start by discussing the different NS  contributions to the  KN light curve in order to analyze the ejecta parameters characterizing the BNS merger event and the subsequent EM emission. As previously obtained, one key aspect is the determination of the peak values in the bolometric light curve as they are related to the ejecta properties. In order to do this we must note that it  is necessary to perform a smoothing, i.e. rebinning, procedure in order to manage the Monte Carlo noise  from the set of simulations we use in our work.

Previous studies assume spherical symmetry although asymmetric ejecta have also been considered \citep{2017ApJvillar}. State-of-the-art calculations \citep{tanaka2020} show that the opacities span the range $\kappa \in[0.1-100]\,\mathrm{cm}^{2} \mathrm{~g}^{-1}$ depending on properties of the ejecta as $\rho$, $T$ and $Y_{\rm e}$. As shown in Fig.~\ref{fig_rhoTrunAB}, models show a toroidal component for $Y_{\rm e}$ appearing clearly where neutron-rich elements concentrate versus a neutron-poor component for low $\theta$ values. The existence of a root mean square angle $\theta_{rms}$ is often quoted, see for example \citet{Radice}, and describes that $\sim75\%$ of the ejecta is confined within $2\theta_{rms}$. Temperatures are approximately isotropic with additional enhancement towards equatorial orientations. These dependencies reflect on the Planck mean opacities and modulate the EM signal we obtain.

Note that the amount of matter ejected in the BNS merger will be only a tiny fraction of the total mass $M_1+M_2$ as shown by detailed NR simulations. Dynamical mass and post-merger disk wind components play a fundamental role in determining the peak magnitudes. In this spirit and considering the fact that the ratio of dynamical to disk mass is less than $\sim 10\%$ for the EoS under discussion we define $M_{dd}=M_{\rm dyn}+M_{\rm disk}$. {This quantity will be relevant for our EoS analysis as it is the matter stripped from the most external layers of the individual NSs during the merger. However, not all this mass will be dinamically ejected from the gravitational potential at the merger site since part of it remains for longer time scales than the measurable change in luminosity. These stripped layers are sensitive to the underlying EoS.} We have verified that, for $q=1$, this mass component  can be fit to a Woods-Saxon distribution of the type $M_{dd}=a(1+\mathrm{exp}(b(M_{\mathrm{chirp}}-c))^{-1}$. More challenging, due to our reduced subset of runs with $q\ne1$, is the description of those BNS mergers with different  individual NS masses. As a tentative fit we use a multiplicative factor for $q\gtrsim 1$ under the functional form $M_{dd}(q)/M_{dd}(1)\sim 1+g(q-1)^h$. We give the approximate fitting parameters for both dependencies in Table \ref{table:fit_mdd_mchirp1}.

\begin{table}[H]
\centering                                      
\resizebox{0.5\textwidth}{!}{\begin{tabular}{c c c c c c c}         
\hline\hline                        
Equation of state & a & b  & c & g & h\\    
\hline                                   
DD2 & 0.3 $\pm$ 0.1  & 12 $\pm$ 6 & 1.2 $\pm$ 0.1 & -0.67 $\pm$ 0.03  & -0.07 $\pm$ 0.02 \\
SFHo & 0.0896 $\pm$ 0.0008 & 168 $\pm$ 5 & 1.1813 $\pm$ 0.0002 & 0.78 $\pm$ 0.03 & 0.08 $\pm$ 0.02  \\
LS220 & 0.19 $\pm$ 0.02 & 30 $\pm$ 7 & 1.18 $\pm$ 0.01& -0.80 $\pm$ 0.07 & 0.03 $\pm$ 0.04 \\
\hline                                           
\end{tabular}}
\caption{Fitting parameters $a,b,c$ of the dynamic plus disk mass, $M_{dd}=M_{\rm dyn}+M_{\rm disk}$ in units of $M_\odot$, as a function of $M_{\rm chirp}$ (for $q\sim 1$),  $M_{\rm dd}(M_{\rm chirp})=a(1+\mathrm{exp}(b(M_{\mathrm{chirp}}-c))^{-1}$, and also $g,h$ describing the $q$ dependence (for $q\gtrsim1$) under the form $M_{dd}(q)/M_{dd}(1)\sim 1+g(q-1)^h$.  We consider runs from \citealt{Radice} and \citealt{nedora}. }  
\label{table:fit_mdd_mchirp1}
 \end{table}

\begin{figure}[ht!]
\centering
  \includegraphics[width=0.95 \linewidth]{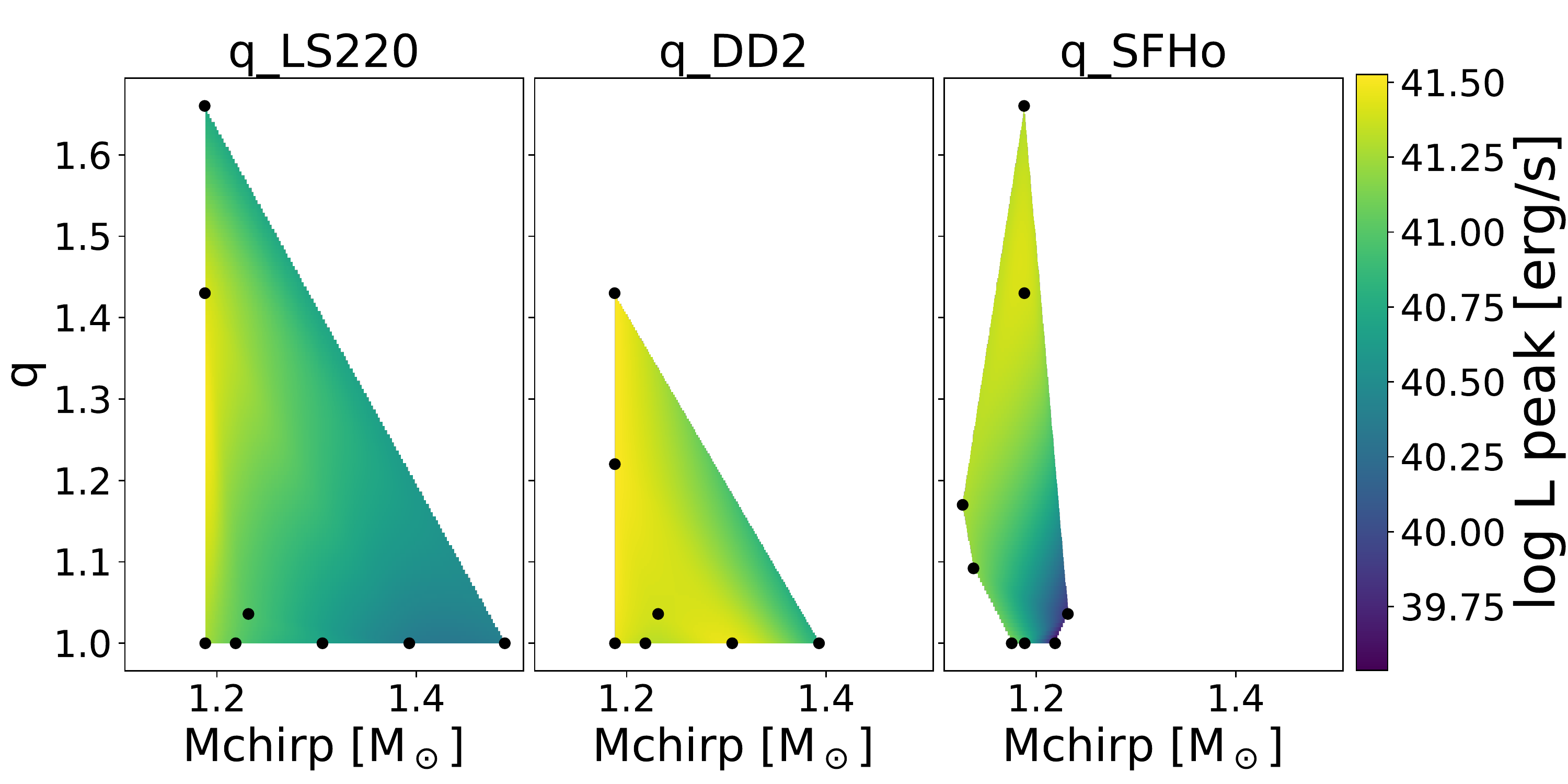}
   \caption{ Logarithm of peak brightness as a function of mass ratio $q$ and chirp mass $M_{\rm chirp}$ for EoS (from left to right) DD2, LS220 and SFHo. Black dots show runs from Table \ref{tab1}. Polar view is selected.}
  \label{figcarlos}
\end{figure}

Figure \ref{figcarlos} shows the extrapolated values of the logarithm of peak brightness as a function of mass ratio $q$ and chirp mass $M_{\rm chirp}$ for EoS (from left to right) DD2, LS220 and SFHo. We select a polar view orientation. We can see how the imprint of the particular EoS along with the BNS realizations will affect the behavior of the peak.

\subsection{KN peak magnitudes and EoS}
According to previous semi-analytical estimates extracted from numerical simulations \citep{Li1998,metzger,Kawaguchi2020}, the luminosity and  peak time for a BNS merger show dependencies on ejecta properties $M_{\rm{ej}}$, $v_{\rm ej}$,   $\kappa_{\rm Planck}\equiv\kappa$, namely the ejected mass, ejected velocity, Planck mean opacity. { In our work, however, we phenomenologically choose to explore the dependence of $L_{\text {peak }}$ on $M_{dd}$, $v_{\rm ej}$, $\kappa$, $Y_e$, due to the usefulness of $M_{dd}$ in relation to the EoS trends}.  We extract trends for $ L_{\text {peak }}$ as $L_{\text {peak}}=L_0\, (M_{dd}/M_\odot)^\alpha \, (v_{\rm ej}/c)^\beta \, (Y_e)^\gamma \, (\kappa/10 \,\rm cm^2 g)^{\delta}$ for edge-on orientation $(\cos \theta=0)$. Given the high-opacity conditions near the peak we fix $\kappa=10$ $\rm cm^2 g$ and $\delta=-0.5$ (see Table \ref{table:Lpkappa}).

\begin{figure}[ht!]
  \includegraphics[width=1.05\linewidth]{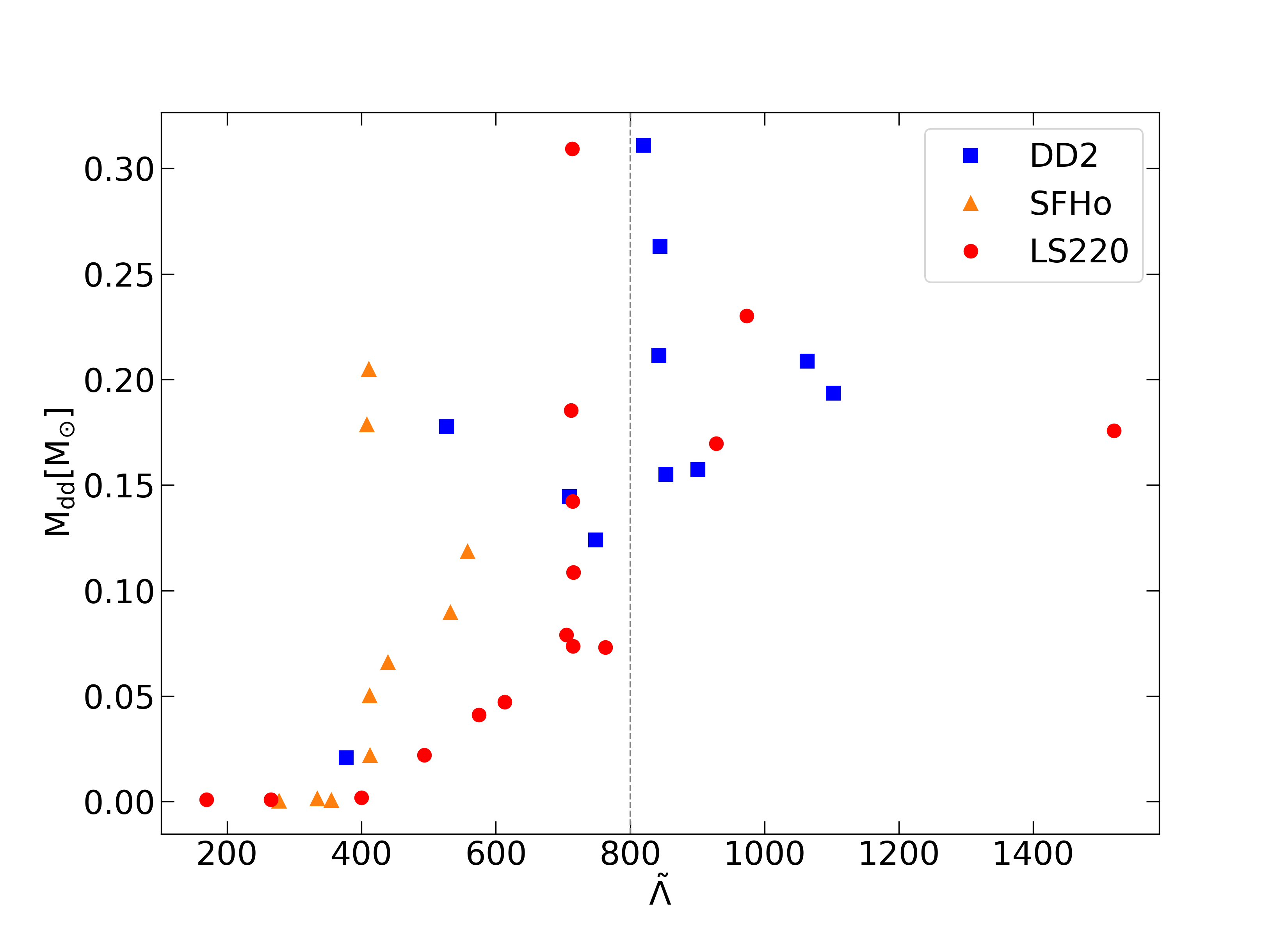}   \caption{Dynamical plus disk mass, $M_{dd}$, as a function of $\mathsf{\tilde{\Lambda}}$ for runs from \citep{Radice} and \citep{nedora}, along with the $\tilde{\Lambda}=800$ limit \citep{abbott2017a}  depicted as a vertical dashed line.} 
  \label{Fig:mchirp_lam}
\end{figure}

Figure \ref{Fig:mchirp_lam} shows the $M_{dd}$ mass for the three EoS under inspection  as a function of $\tilde{\Lambda}$. The latter is available as a  GW measured quantity from a BNS event and from GW170817 there is an upper bound set at $\tilde{\Lambda} \lesssim 800$ \citep{abbott2017a} that we show as a vertical line. We notice that some of the DD2, LS220 instances we have considered do not fulfill this constraint. Let us highlight at this point that only for a subset of these, the runs appearing in Table~\ref{tab1}, we have obtained KN spectra and light  curves. 

\begin{table}[t!]
\centering                                      
\resizebox{0.5\textwidth}{!}{\begin{tabular}{c c c c c }         
\hline\hline                        
EoS & $\mathsf{log (L_{0} [erg/s])}$ & $\mathsf{\alpha}$ & $\mathsf{\beta}$  & $\mathsf{\gamma}$  \\    
\hline                                   
 DD2    & 43.1 $\pm$ 0.4 & 1.0 $\pm$ 0.1 & 1.9 $\pm$ 0.5 & -0.6 $\pm$ 0.2  \\      
 SFHo  & 41.0  $\pm$ 0.4 &  0.7  $\pm$ 0.1      & 0.9      $\pm$ 0.6 &  -1.8 $\pm$ 0.4 \\
 LS220 & 42.0  $\pm$ 1.5 &  0.3  $\pm$ 0.2      & 1.3      $\pm$ 1.9 &  -0.44 $\pm$ 0.43 \\
\hline                                      \label{tab:mdd_lamtilde}    
\end{tabular}}
\caption{Fit of peak luminosity $L_{\text {peak }}$ as $L_{\text {peak}}=L_{1d}\, (M_{dd}/M_\odot)^\alpha \, (v_{ej}/c)^\beta \, (Y_e)^\gamma \, (\kappa/10 \,cm^2 g)^{\delta}$ for selected runs as a function of ejecta parameters $M_{\mathrm{ej}}, v_{\rm ej}, Y_e$. For peak conditions we fix $\kappa=10$ $\rm cm^2 g$ and $\delta=-0.5$. Equatorial view is selected. See  text for details.}  \label{table:Lpkappa}
\end{table}

\begin{figure}[ht!]
  \includegraphics[width=1.05\linewidth]{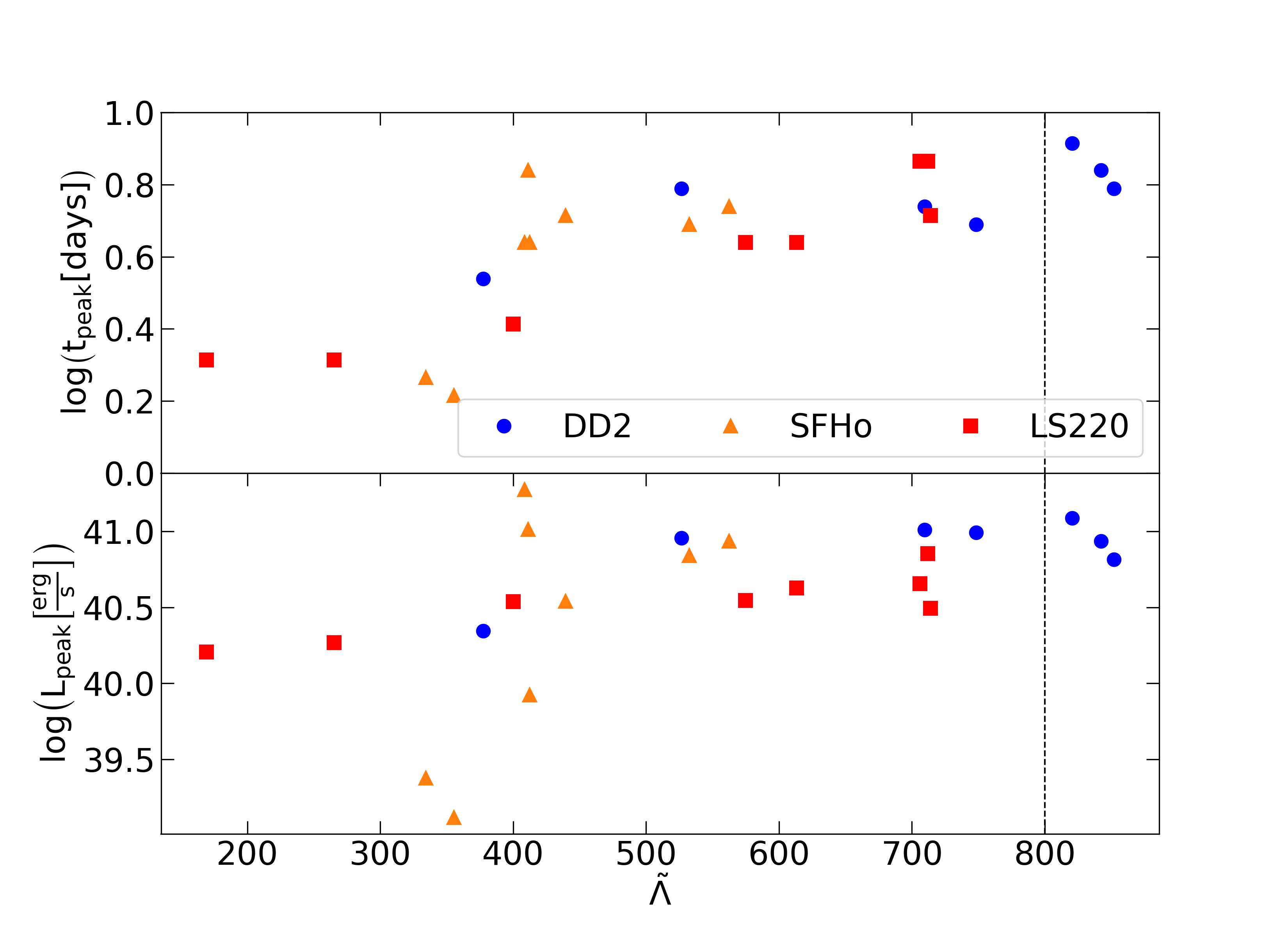}\caption{Time (upper panel) and luminosity (lower panel) at the peak for equatorial orientation and for all runs in Table \ref{tab1}. We  also indicate the upper limit $\mathrm{\tilde{\Lambda}=800}$ \citep{abbott2017a}.}
  \label{Fig:peak_lam}
\end{figure}

Interestingly, the exclusion region set by the $\tilde{\Lambda} \lesssim 800$ upper bound on $M_{dd}$ has relevant consequences on the peak log luminosity and time as  shown in Fig \ref{Fig:peak_lam}. We select equatorial orientation. Although the set of runs we consider in Table \ref{tab1} is reduced it seems to indicate the presence of limiting values for peak magnitudes. If such a case is confirmed by larger statistical samples it would be a genuine limit for the intrinsic magnitudes of KNe. From our plot we derive log $L_{peak} [\rm erg/s]\lesssim 41$ and log $t_{peak} [\rm day]\lesssim 0.9$. Note that the previous data points and bounds refer to an equatorial inclination, when considering a polar observer we obtain an enhanced luminosity yielding a bounding value log $L_{peak} [\rm erg/s]\lesssim 41.3$.
Additional investigations are needed regarding the robustness of our finding and a careful reexamination of the caveats we have used in the simulations.
\begin{figure}[ht!]
  \includegraphics[width=1.05\linewidth]{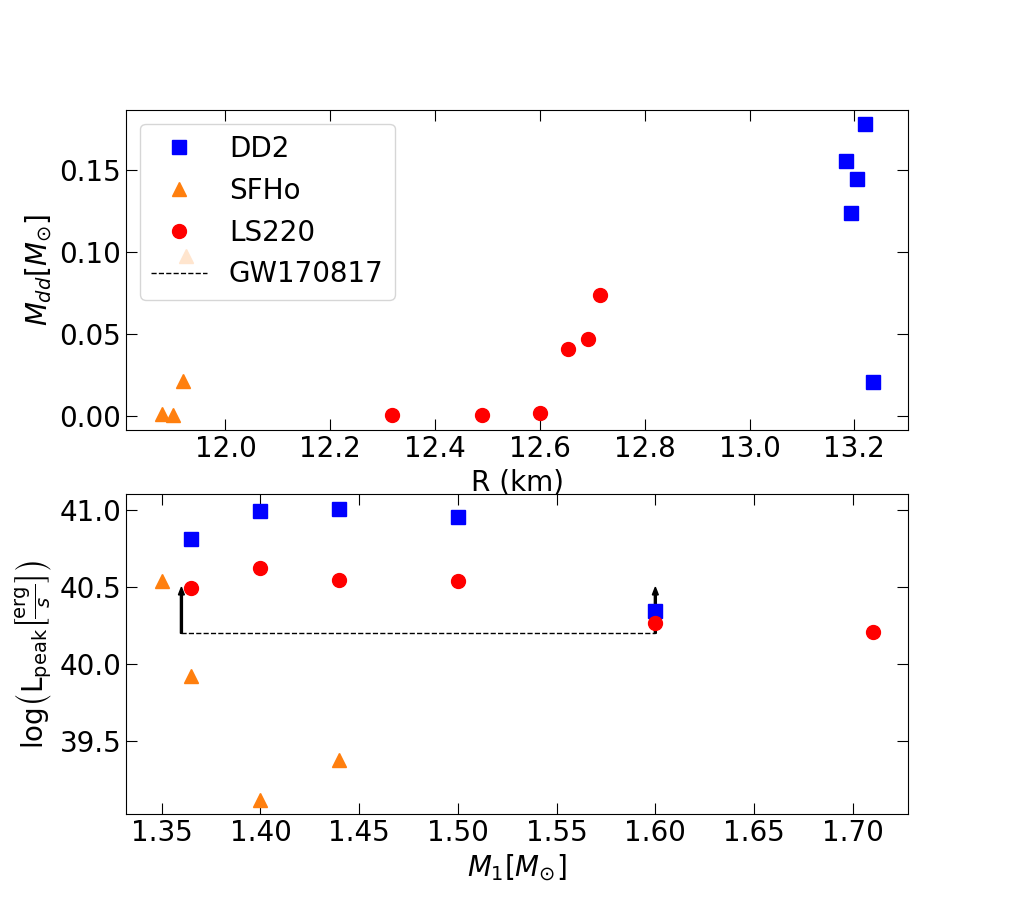}\caption{(Top panel) $M_{dd}$ as a function of NS radius from NR runs due to \citealt{Radice} and \citealt{nedora} having $q \simeq 1$. (Bottom panel)  $\mathsf{log\left(L_{peak}[\rm erg/s]\right)}$ for an equatorial orientation as a function of primary mass $\mathsf{M_1}$, for runs in Table \ref{tab1} also having $q\simeq1$.We also indicate a conservative lower limit estimate of equatorial peak luminosity for a GW170817-like transient.}
  \label{fig:Lpeak_M1}
\end{figure}


In order to better constrain the  allowed regions on the Mass-Radius relation and thus the EoS (see Fig. \ref{Fig:q_mc_m1}), we select runs from \cite{nedora} and \cite{Radice} with $q\simeq 1$. Top panel in 
 Fig. \ref{fig:Lpeak_M1} shows $M_{dd}$ as a function of the radius for DD2, LS220 and SFHo EoS. We can clearly see that for radii smaller than a given value (EoS dependent) the stripped mass dramatically plunges to vanishing values. For DD2 the NR cases are in the nearly vertical part of the $M/R$ curve, thus this trend is not seen clearly, although radii smaller than $\sim 13.2$ seem to exclude observable KNe. Thus we do not expect having KNe being observed for NSs with compactness beyond a limiting value as dictated by these radii. Our results agree with those for AT 2017gfo appearing in \cite{Breschi} obtained from Bayesian analysis and semi-analytical models except for the SFHo EoS, since in our analysis for that specific EoS a BNS with $M=1.4\,M_\odot$ and $q\sim 1$ luminosity peaks below  experimental values. Additional correlations to central densities in TOV solutions have been explored \citep{rhocen}. It is worth noting at this point that it is likely that there will be an additional EoS dependence associated to the BNS redshift that must be carefully analyzed, see \citep{Ghosh22}, as it can impact the measurement of $H_0$.
 
 In the lower panel of Fig. \ref{fig:Lpeak_M1} and for the runs from Table \ref{tab1} with  $q\simeq 1$ where KN light curves are simulated we show $\mathsf{log\left(L_{peak}[\rm erg/s]\right)}$ for an equatorial orientation as a function of the NS mass in the BNS, $\mathsf{M_1}$. In the same figure we indicate the deduced equatorial lower limit for the peak luminosity for AT 2017gfo (in dashed line). {One can observe that for LS220 and DD2 EoS, the luminosity threshold for AT 2017gfo is reached. 
 The only effect that drives larger peak luminosities between the two sets of values is the radius of the individual NSs, i.e. the compactness. Let us remind here that $C_{\rm DD2}<C_{\rm LS220}$ for the stiffer DD2 EoS versus the softer LS220 EoS. Even though the effect seen here is only qualitative, due to poor statistics induced by our reduced set of simulated models, being able to have a library of NR simulated runs with different primary NS masses will allow us to isolate the effect of the EoS in the KNe luminosities. For SFHo EoS the high NS compactness predicted reduces the amount of ejected material and so the luminosity of the transient indicating that it seems not well suited to describe this KN. Additional analysis with improved statistics will be able to provide stronger constraints on the EoS. }

\section{KN light curves}
\label{ref:KNlc}
After the BNS merger, the KN emission follows due to a combination of signals characterized by how nucleosynthesis proceeds. A red component is associated with a lanthanide-rich outflow, and a blue  component with the lanthanide-poor counterpart. The high opacity of a lanthanide-rich (heavy $r$-process) ejecta delays and dims the light curve peak, and its large density of lines at optical wavelengths pushes the emission to redder wavelengths. 
Light $r$-process compositions will instead have a lower opacity and the emission associated with these outflows will have a faster rise and brighter luminosity peak with fluxes concentrated at shorter/bluer wavelengths. Alternative scenarios to explain the blue to red spectral evolution,  such as dust formed by the KN, which  extinguishes the light from the blue component and subsequently re-emits it at NIR wavelengths, have also been suggested \citep{Takami}. However, for AT 2017gfo this could be ruled out \citep{Gall17}.  

As the envelope becomes more transparent due to the expansion while the radioactive power decreases, the luminosity initially rises to a maximum (peak) before declining. A variety of KN models are available ranging from simple (semi-analytical), spherical, radioactive heating parametrized by a power-law and black-body emission \citep{Li1998,Kawaguchi2016,Metzger2015,metzger} to more refined ones that are nonspherical, with detailed modelling of radioactive heating and radiative transport  \citep[e.g.][]{Kasen2017,Bulla2019,Kawaguchi2020,Korobkin2021}.

The radioactive decay of freshly synthesized $r-$process nuclei in the expanding ejecta at relativistic velocities powers a quasi-thermal emission responsible for the KN. The peak flux ${F}_{\lambda,\rm peak}$  and peak time $t_{\lambda,\rm peak}$ in a given spectral band depend both on intrinsic (ejected mass, expansion velocity, composition) and extrinsic (viewing angle, distance, reddening) parameters. The 23 runs considered in this work (see Section~\ref{sec:modelsandEOS}  
and Table~\ref{tab1}) differ in terms of the ejecta properties like mass, $M_{\rm ej}$, average velocity, $v_{\rm ej}$, and average electron fraction, $Y_{\rm e}$.  The latter, in particular, determines the photon opacity for a given wavelength of the broadband radiation, which can substantially differ from the opacity of iron group elements in the presence of a significant fraction of moderate ($A=90-140$) or heavy elements \citep{tanaka2020}. It is still unclear if NS mergers are the main source of $r-$process elements \citep{Siegel2019}, although high-opacity lanthanides have been inferred in GW170817 (see \citealt{Ji} and references therein).

Opacities from \cite{tanaka2020} used in this work are underestimated at $t\lesssim0.5$ d, due to the lack of atomic data at high ionization states, typical for these early epochs. Therefore, we focus our discussion on times $t\gtrsim0.5$\,d. In the same way the decline of the light curves shows two regimes, before and after the peak that are numerically determined by analyzing where  the rapid decline starts. Near the peak, the structure of the luminosity is not trivial and displays some variability due to underlying dependencies on $q/M_{\rm chirp}$ and EoS physics. In Fig. \ref{Fig:lightcurves} we show the behaviour near the luminosity peak by plotting the spread in the logarithm of the luminosity as obtained frpeom the simulations with respect to a power law $L\sim t^\gamma$ law, $\Delta L$ in units of erg/s, as a function of time, in days, for different orientations. The closer to the power law $L\sim t^\gamma$ before the peak, the darker bluish the color, while the larger deviation is coded in reddish colors. The peak is denoted by a black dot. Each particular run we consider shows a distinctive feature depending not only on the EoS but also on the ejecta parameters. Runs with the same EoS show a big spread near the peak as $q$ and $M_{\rm wind}$ grows. Throughout the manuscript we label this point in the KN light curve of our analysis as peak brightness.

\begin{figure}
  \includegraphics[width=0.99\linewidth]{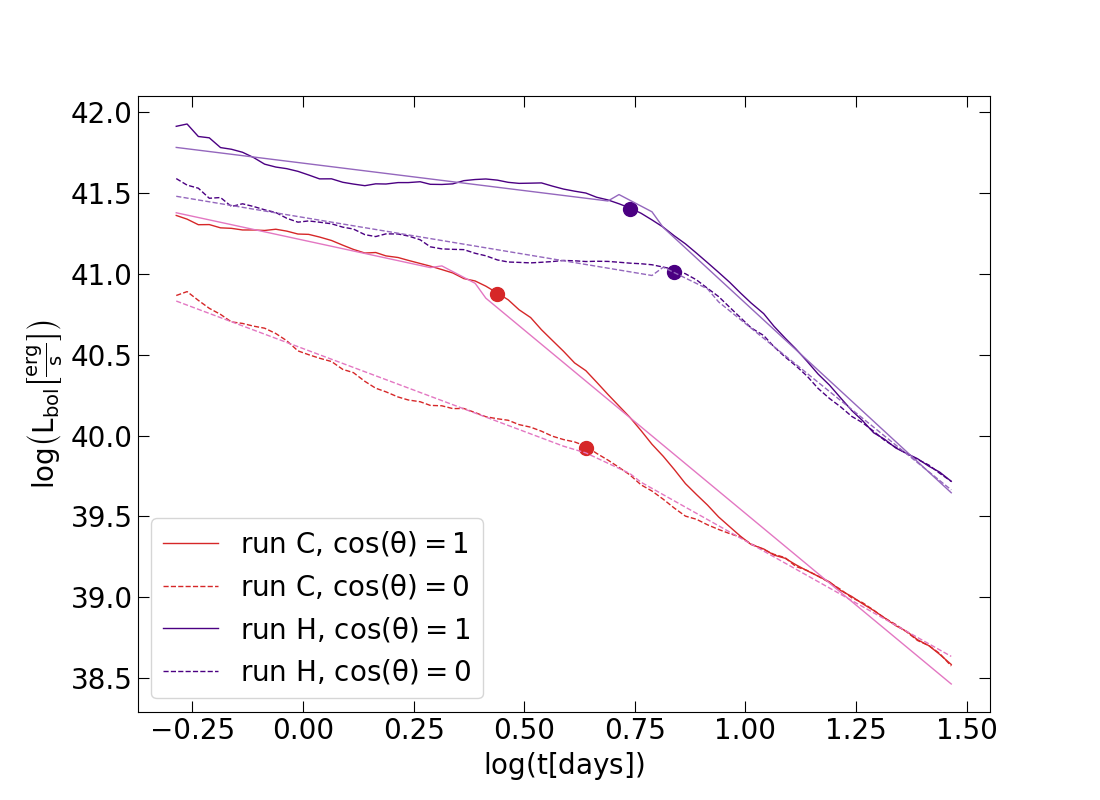}
   \caption{Log $\rm{L[erg/s]}$ as a function of log $\rm{t[d]}$ comparing simulated models versus fits for runs C and H using SFHo EoS, see Table \ref{tab1}, for polar and equatorial orientations.The inflexion point signals the change in the decay law as appears in Table \ref{table:chfit}. 
   }
 \label{fit_CH}
\end{figure}
%
From semi-analytic estimates in \cite{2017metzger}, the time-dependent KN bolometric luminosity, assuming homogeneous and homologous ejecta has a functional form 
\begin{equation}
L(t)\simeq L_{1d} \left\{\begin{array}{l}
\left(\frac{t}{1 \mathrm{~d}}\right)^{-\alpha+1}, \quad t \leqslant t_{peak} \\
\left(\frac{t}{1 \mathrm{~d}}\right)^{-\alpha}, \quad t>t_{peak}
\end{array}\right.
\end{equation}
where $\alpha=1.3$. The peak luminosity $L_{\text {peak }}$ and time $t_{\text {peak }}$ are determined \citep{Kawaguchi2020}  roughly to be equivalent to the instantaneous rate at which radioactive decay is heating the ejecta. The EoS dependence will appear in these luminosity curves built in an indirect way in the ejecta parameters and composition.

\begin{figure*}[ht!]
\centering
\includegraphics[scale=0.6]{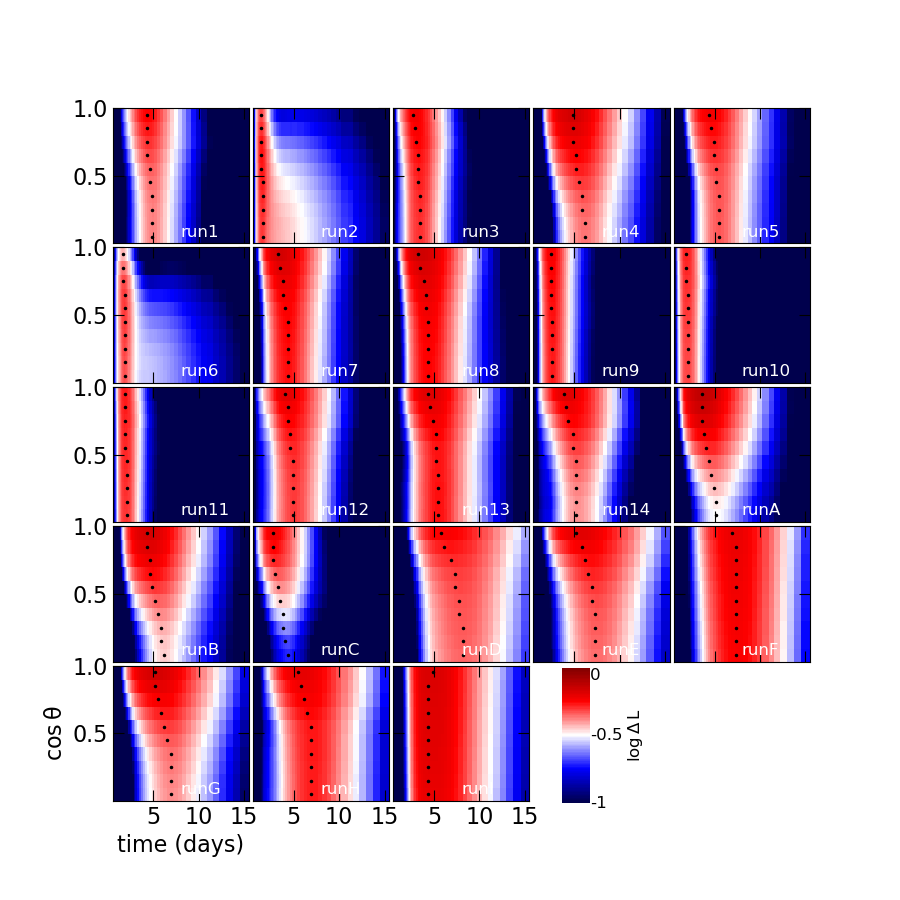}
\caption{Each panel represents a model run from Table~\ref{tab1} and displays the departure of the logarithm of the luminosity from a power law $L\sim t^\gamma$  near its peak as a function of the orientation, $\cos\theta$, and time (in days) elapsed since merger. Black dots mark the location of the peak derived from our analysis. See text for details.}
\label{Fig:lightcurves}
\end{figure*}

As mentioned, the material ejected during and after the merger of two NSs (or a NS and a black hole) is strongly spatially asymmetric and characterized by multiple ejecta components with different geometries and properties, see \cite{Nakar2020} for a recent review. As a result, the KN signal is found to be strongly dependent on the observer viewing angle \citep{Bulla2019,Darbha2020,Kawaguchi2020,Korobkin2021}, with orientations in the merger plane (equatorial or edge-on inclinations) typically fainter and redder than orientations on the jet axis (polar or face-on inclinations). As a result, we end up with a viewing-angle dependence of the signal  that offers a way to constrain the inclination angle from the KN. 
In order to qualify the impact of the inclination in the measured brightness decline, we have considered as examples runs C and H with SFHo EoS from \citet{nedora}, see Table~\ref{tab1}. Both runs have a chirp mass $\mathcal{M}=1.188 M_{\odot}$, but different mass ratios $q=1, 1.43$ respectively. Figure~\ref{fit_CH} shows log L in units of erg/s as a function of log $t$ in days for those mentioned simulations along with fitting functions. Solid (dashed) lines depict polar (equatorial) orientation. We can see that realistic models such as these are well represented by log-log laws. Table \ref{table:chfit} compiles the actual parameters describing the region before (after) the peak with corresponding values labeled as bp (ap).

\begin{figure}[ht!]
  \includegraphics[width=1.05\linewidth]{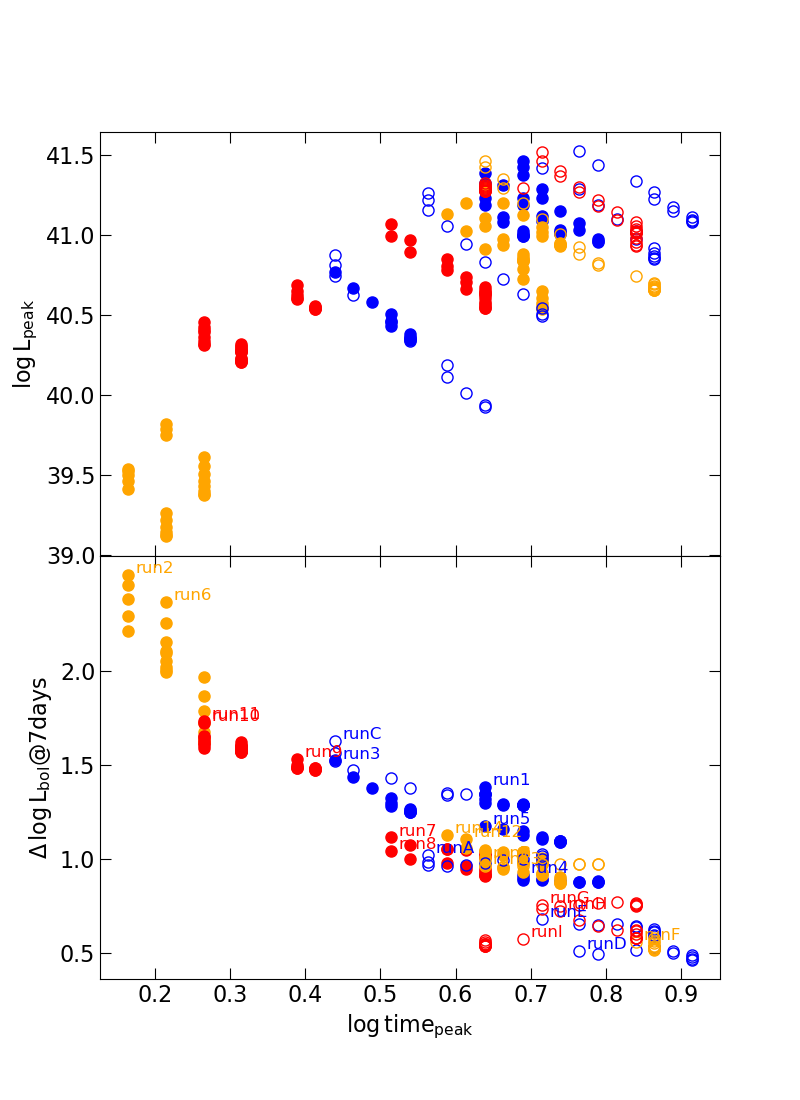}
   \caption{ Luminosity in erg/s of peak (top panel) and seven day gradient (bottom panel) in log bolometric luminosity as a function of peak time in days.  
   Full and empty symbols, only made explicit in the bottom panel for the sake of clarity, correspond to models listed in Table~\ref{tab1} from \citet{Radice} and \citet{nedora}, respectively. For each run, orientations are depicted by the dot series. LS220, SFHo and DD2 EoS are shown in red, orange and blue colors, respectively.}
  \label{fig_runs67.png}
\end{figure}

\begin{table}[H]
\centering                                      
\resizebox{0.5\textwidth}{!}{\begin{tabular}{c c c c c c}          
\hline\hline                        
Model & $\cos \theta$ & $\beta_{bp}$ & $\alpha_{bp}$  & $\beta_{ap}$ & $\alpha_{ap}$ \\    
\hline                                   
  C &  1 & 41.21 $\pm$ 0.01 & -0.59 $\pm$ 0.02 & 41.798 $\pm$ 0.05 &  -2.27 $\pm$ 0.05 \\      
  C & 0 & 40.54 $\pm$ 0.01& -1.03 $\pm$ 0.03      & 40.89 $\pm$ 0.02 & -1.54 $\pm$ 0.02 \\
  H &  1 & 41.69 $\pm$ 0.01 & -0.34 $\pm$ 0.03  & 43.30 $\pm$ 0.04 & -2.52 $\pm$ 0.03 \\
  H &  0 & 41.35 $\pm$ 0.01 & -0.47 $\pm$ 0.02     &  42.90 $\pm$ 0.03 & -2.21 $\pm$ 0.03 \\
\hline                                       
\end{tabular}}

\caption{Fit parameters for $\rm{log \,L=\beta+\alpha log\, t}$ (L in erg/s and t in days) in the bolometric luminosity before/after the peak (bp/ap) for polar and equatorial orientations. We use runs C and H with SFHo EoS, see Table \ref{tab1}. }   
\label{table:chfit}
\end{table}
As we can observe, there is a clear difference in the polar to equatorial orientation between the fit parameters $\beta$, and $\alpha$, according to the law  log $L=\beta+\alpha \,\rm{log} t$ for the bolometric luminosity. We have verified that this also applies to the peak values, and as a general trend this behaviour is seen for all the runs we consider. Figure \ref{fig_runs67.png} illustrates this. In the top panel we show the logarithm of the luminosity of the peak as a function of logarithm of peak time for the models we consider in this work. As usual L is expressed in erg/s and $t_{peak}$ in days. In the bottom panel we show the 7-day variation in the logarithm of the bolometric luminosity of peak as a function of logarithm of peak time. LS220, SFHo and DD2 EoS are shown in red, orange and blue colors, respectively. For a given run, full and empty symbols represent different $\cos\theta$ orientations correspond to models listed in Table \ref{tab1} from \citet{Radice} and \citet{nedora}, respectively. The plot in log-log scale shows clear correlations among peak luminosity and time of the peak: the brighter the KN is, the longer it takes to peak and the smoother the change in magnitude is. For each run the dot series depicts the angular modulation of the peak value. 

\begin{figure}[ht!]
  \includegraphics[width=1.09\linewidth]{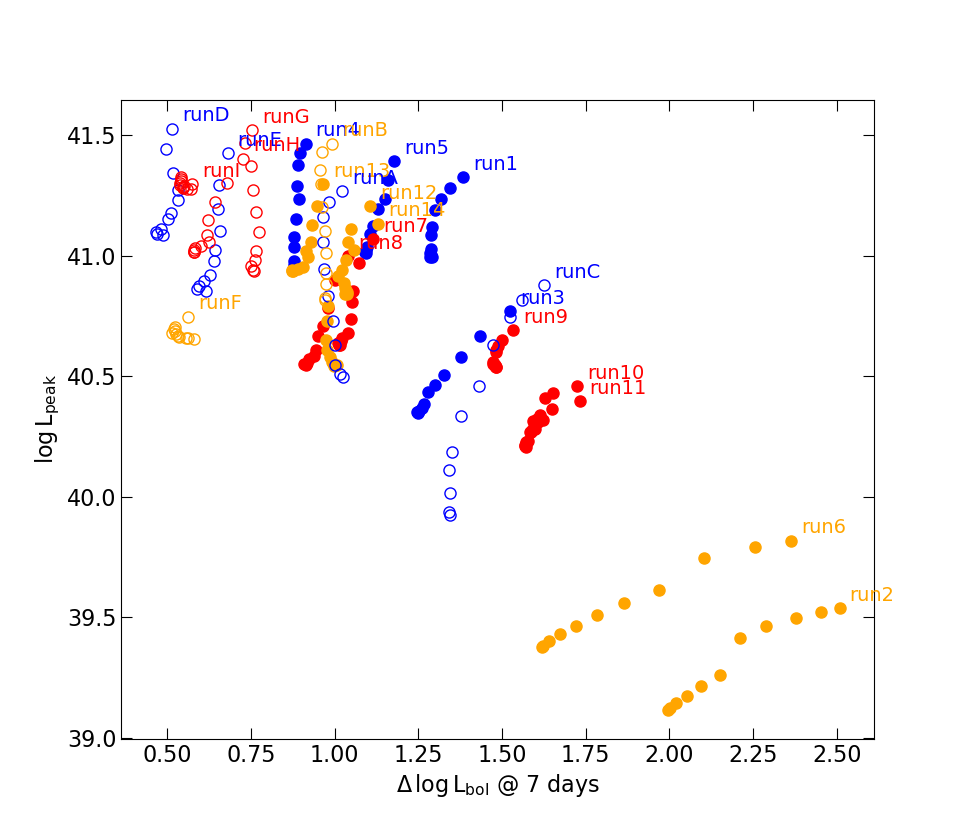}
   \caption{Luminosity in erg/s as a function of seven day gradient in log bolometric luminosity. Full and empty symbols correspond to models listed in Table \ref{tab1} from \citet{Radice} and \citet{nedora}, respectively. For each run, the orientations are depicted by the dot series. LS220, SFHo and DD2 EoS are shown in red, orange and blue colors, respectively.}
  \label{Fig:gradient7_time}
\end{figure}

In addition, Fig. \ref{Fig:gradient7_time} shows the  log peak luminosity (in erg/s) as a function of seven day gradient in log bolometric luminosity (in erg/s). We can see that the fainter they are the larger decay  in bolometric luminosity after a week they experience.
As seen from Figs. \ref{fig_runs67.png} and \ref{Fig:gradient7_time}, there is a clear modulation from observing angle $\theta$. In order to quantitatively determine its impact we have considered the models in Table~\ref{tab1} corresponding to a common EoS and fit the angular dependence as obtained from the 11 orientations for each simulated run.  Table \ref{table:fit_lpcos} shows the results of a fit to the peak value log $L_{\rm peak}(x) = c_0+c_1 {\rm \arctan}(c_2(x-c_3))$ with $x=\cos \theta$. At high inclinations (low $x$ values), ${\rm arctan} (x)\simeq x-\frac{x^{3}}{3}$. We find that, generally speaking, a function of this type describes satisfactorily the angular modulation of luminosities. 

\begin{figure}
  \includegraphics[width=1.08\linewidth]{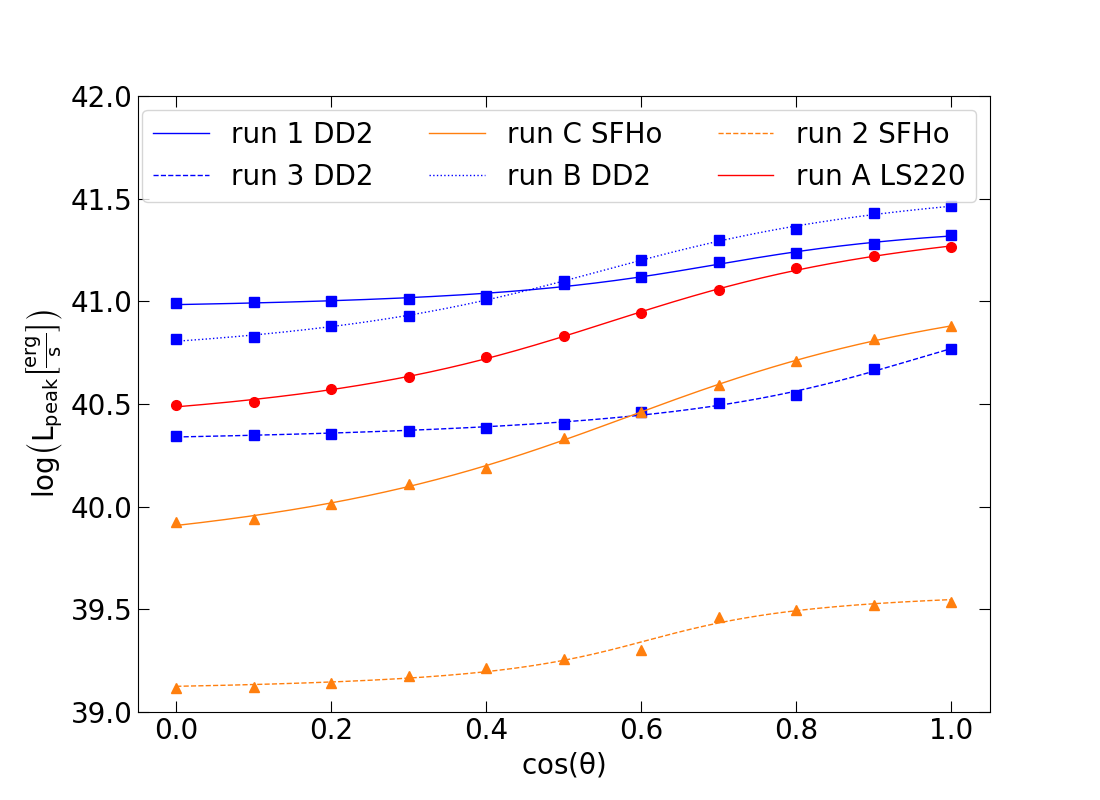}
   \caption{Peak bolometric luminosity as a function of $\cos \theta$ and fitting functions. LS220, SFHo and DD2 EoS are shown in red, orange and blue colors, respectively. Runs A, B and C have a fixed  ${M}_{\rm chirp} = 1.118 M_{\odot}$. Runs 1, 2 have  ${M}_{\rm chirp} = 1.22 M_{\odot}$ and Run 3 has ${M}_{\rm chirp} = 1.39M_{\odot}$. A fixed $q=1$ value is used.}
  \label{figpeak_cos1}
\end{figure}

\begin{table}[ht!]
\centering                                      
\resizebox{0.5\textwidth}{!}{\begin{tabular}{c c c c c c }         
\hline\hline                        
Model & $q/M_{\rm chirp}[M_\odot]$ & $c_0$ & $c_1$  & $c_2$ & $c_3$ \\    
\hline                                   
run 1 DD2 & 1/1.22 & 41.18 $\pm$ 0.01 & 0.16 $\pm$ 0.02 & 4.1 $\pm$ 0.7 & 0.69 $\pm$ 0.02 \\ 
run 3 DD2 & 1/1.39 & 40.7 $\pm$ 0.1 & 0.3 $\pm$ 0.1 & 3.9 $\pm$ 0.6 & 0.95 $\pm$ 0.08 \\ 
run B DD2 & 1/1.188 & 41.15 $\pm$ 0.01 & 0.33 $\pm$ 0.02 & 3.0 $\pm$ 0.3 & 0.55 $\pm$ 0.01 \\ 
run C SFHo & 1/1.188 & 40.44 $\pm$ 0.01 & 0.54 $\pm$ 0.03 & 2.6 $\pm$ 0.3 & 0.58 $\pm$ 0.01 \\ 
run 2 SFHo & 1/1.22 & 39.35 $\pm$ 0.01 & 0.17 $\pm$ 0.02 & 5.9 $\pm$ 1.5 & 0.61 $\pm$ 0.02 \\ 
run A LS220 & 1/1.188 & 40.90 $\pm$ 0.01 & 0.40 $\pm$ 0.02 & 3.0 $\pm$ 0.2 & 0.56 $\pm$ 0.01 \\ 
\hline                                           
\end{tabular}}
\caption{Fit of the expected peak bolometric luminosity (in erg/s) as a function of $x=\cos\theta$,  log $L_{\rm peak}(x) = c_0+c_1 {\rm \arctan}(c_2(x-c_3))$. We consider runs from Table~\ref{tab1} with selected values of $q$ and $M_{\rm chirp}$.}  
\label{table:fit_lpcos}
\end{table}
This is illustrated in  Fig. \ref{figpeak_cos1}, that shows the logarithm of peak of the luminosity (in erg/s) as a function of  $\cos \theta$ for selected values of mass ratio $q=1$ and $M_{\rm chirp}$. Runs A, B and C have a fixed  ${M}_{\rm chirp} = 1.118\, M_{\odot}$, runs 1, 2 have  ${M}_{\rm chirp} = 1.22 \,M_{\odot}$ while run 3 has ${M}_{\rm chirp} = 1.39\,M_{\odot}$. Note that  $(q,M_{\rm chirp})$ will be provided by analysis following a positive detection from GW detectors. In order to obtain coherent underlying trends we fix the hadronic EoS. LS220, SFHo and DD2 EoS are shown in red, orange and blue colors, respectively. The actual values of the fitting parameters are provided in Table \ref{table:fit_lpcos}.

\section{KN spectro-photometry and $H_{0}$ determination}
\label{H0}


In this section, we study the observational prospects, and Hubble constant determination, by using the new MAAT IFU on the GTC from the potential KNe emission simulated for a sample of NS mergers described above, and listed in Table \ref{tab1}. 

\begin{figure}
    \centering
    \includegraphics[width=0.96\linewidth]{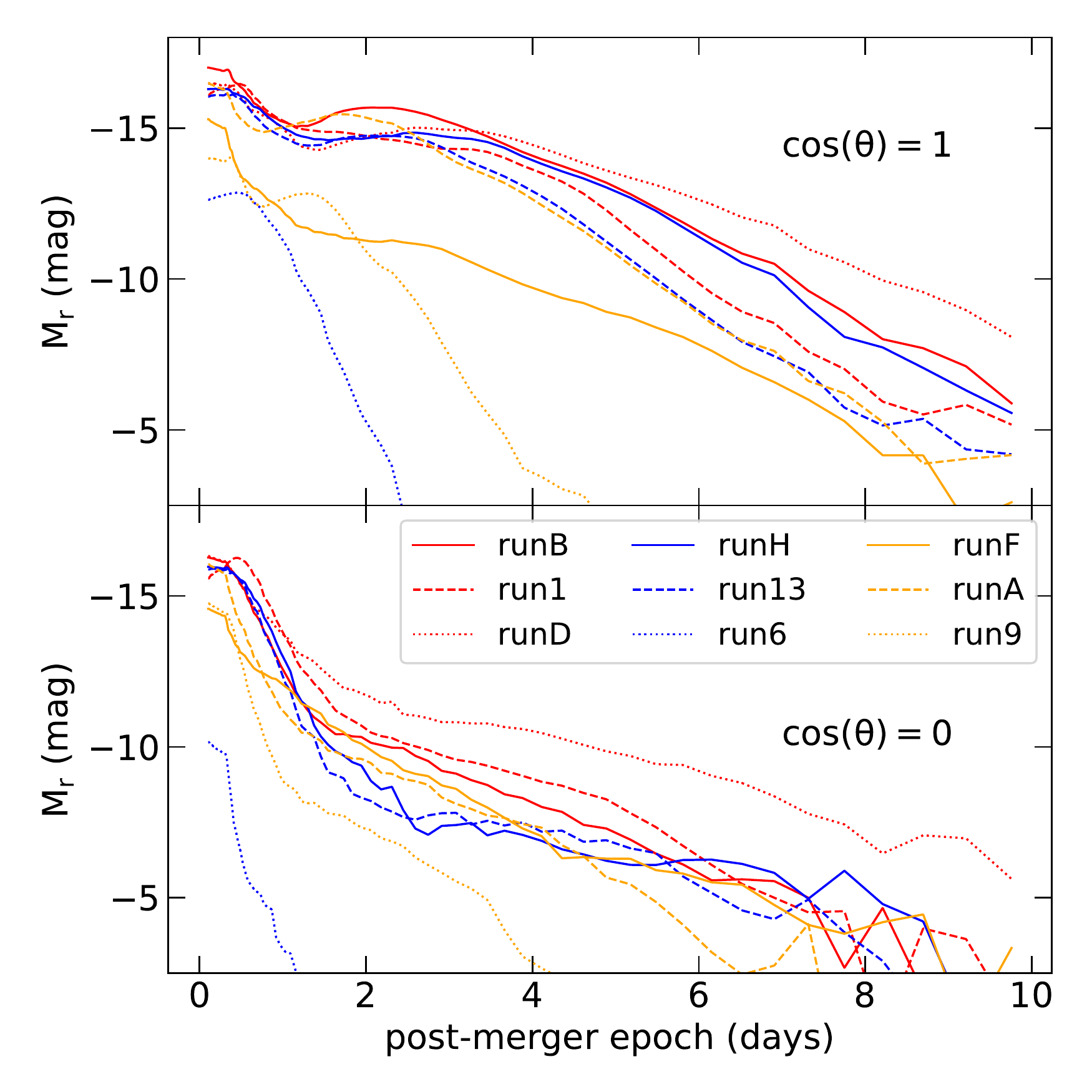}
    \caption{Simulated $r$-band photometry plotted for the KNe models shown in the legend of the figure. In the top panel, we see the results for polar angles ($\theta=0^{\circ}$), and the equatorial viewing angles in the bottom ($\theta=90^{\circ}$). The line style distinguishes between light curves rising from different equation-of-states: DD2 (solid), SFHo (dashed), and LS220 (dotted).}
    \label{fig:absmag_lcs}
\end{figure}

Figure \ref{fig:absmag_lcs} shows the absolute magnitude per post-merger epoch in $r$ band for some of the KNe types that we are considering in our study. This plot shows the drop in observed magnitude when the KN is oriented towards the merger plane of the neutron stars. 
To study the detectability and get an idea of what type of KNe we will be able to observe with MAAT, we explore the dependencies of the S/N   with distance, inclination angle and intrinsic physics of the KNe. 



\begin{figure*}[ht!]
    \centering
    \includegraphics[width=.9\textwidth]{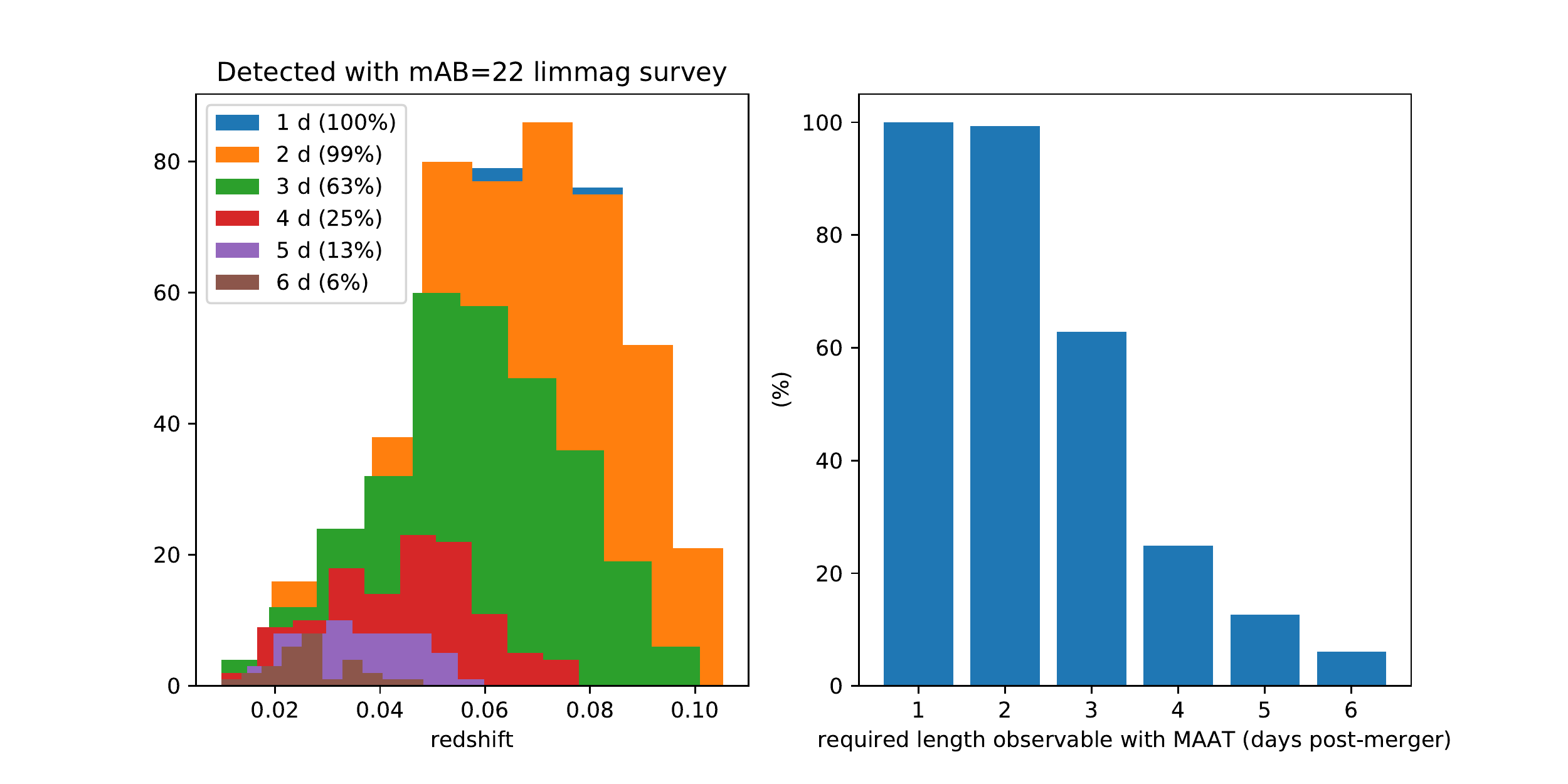}
    \caption{(Left panel) Redshift distribution on simulated KNe detected by a generic optical survey with a nominal depth of $\sim$22 mag, and each color shows the distribution dependence on the detectability per redshift with the start time of the observations. The numbers in parenthesis present the ratio at which the detectability decays with time. (Right panel) Ratio of KNe discovered by a 22 mag depth optical survey that is detectable by MAAT per post-merger epoch. MAAT would be able to follow up on most of the KNe, if not all, that would appear on the observable area. For this computation, a model similar to AT 2017gfo KN was used to distribute inclination angles uniformly in cos($\theta$).}
    \label{fig:m22survey}
\end{figure*}

\subsection{Simulating KN spectro-photometry observations}

We have built a pipeline to generate realistic KN spectro-photometric data as observed with MAAT, and perform an end-to-end analysis of the spectro-photometric light curves. The input of the pipeline consists of templates to model the theoretical flux of the source of interest and provide the relevant information for the MAAT exposure-time-calculator (ETC). The ETC provides the output S/N (exposure time) as a function of wavelength for a requested exposure time (S/N), taking into account the transmission through the atmosphere and the telescope and instrument (OSIRIS+MAAT) optics. 
It works for the whole suite of OSIRIS gratings and for an optional range of source geometries. The complete code package and instructions are available on the MAAT website\footnote{http://maat.iaa.es/maat\_etc.html}. The ETC output includes the observed simulated spectra in physical units (erg/cm$^{2}$/s/$\rm \AA$). We convolve the spectrum with the standard $gri$ SDSS filters transmission curves to extract also synthetic photometry.

We plan to use the MAAT IFU with the OSIRIS R1000B and R1000R grisms to follow-up KNe alerts from the upcoming GW detector runs or after a potential identification by imaging surveys, as described in Section \ref{sec:obs}. We expect to be very sensitive to observe faint KNe, over the $3600-10000$ $\AA$ spectral range, down to $\sim$22 mag with 1800s exposures per grism, see also Fig. \ref{fig:m22survey}. 
Our aim consists of showing the potential of MAAT observations to set meaningful constraints on two of the most desirable research fields related to compact binary mergers and their electromagnetic counterpart: the study of the EoS (see Sec. \ref{sec:modelsandEOS}), and the potential contribution to cosmology as by improving the Hubble constant  estimation from standard sirens in the local universe \citep{Holz2005}.

For population studies and long-term predictions, it is required to collect a significant statistical sample of KN data. However, only one KN has been successfully identified to date from a BNS merger (GW170817-AT 2017gfo). 

We aim to use the standard sirens method from GW data together with prior from the KN data as an independent constraint on the inclination of the system and improve the derived $H_0$ uncertainties. 
We perform the analysis with synthetic MAAT data for the only two BNS mergers detected to date with GW detectors: GW170817 and GW190425.

{In this section, we present synthetic data of follow up KNe observations with MAAT. Therefore, we consider an exposure time of 1800 seconds ($0.5$ hr) with grims R1000B and R1000R. Dark moon, 1 airmass and a seeing of $0.8"$ is assumed for all the simulated data in this analysis.}

\subsubsection{GW170817 KN light curves}\label{subsec:17gfoLC}

\begin{figure}
    \centering
    \includegraphics[width=.96\linewidth]{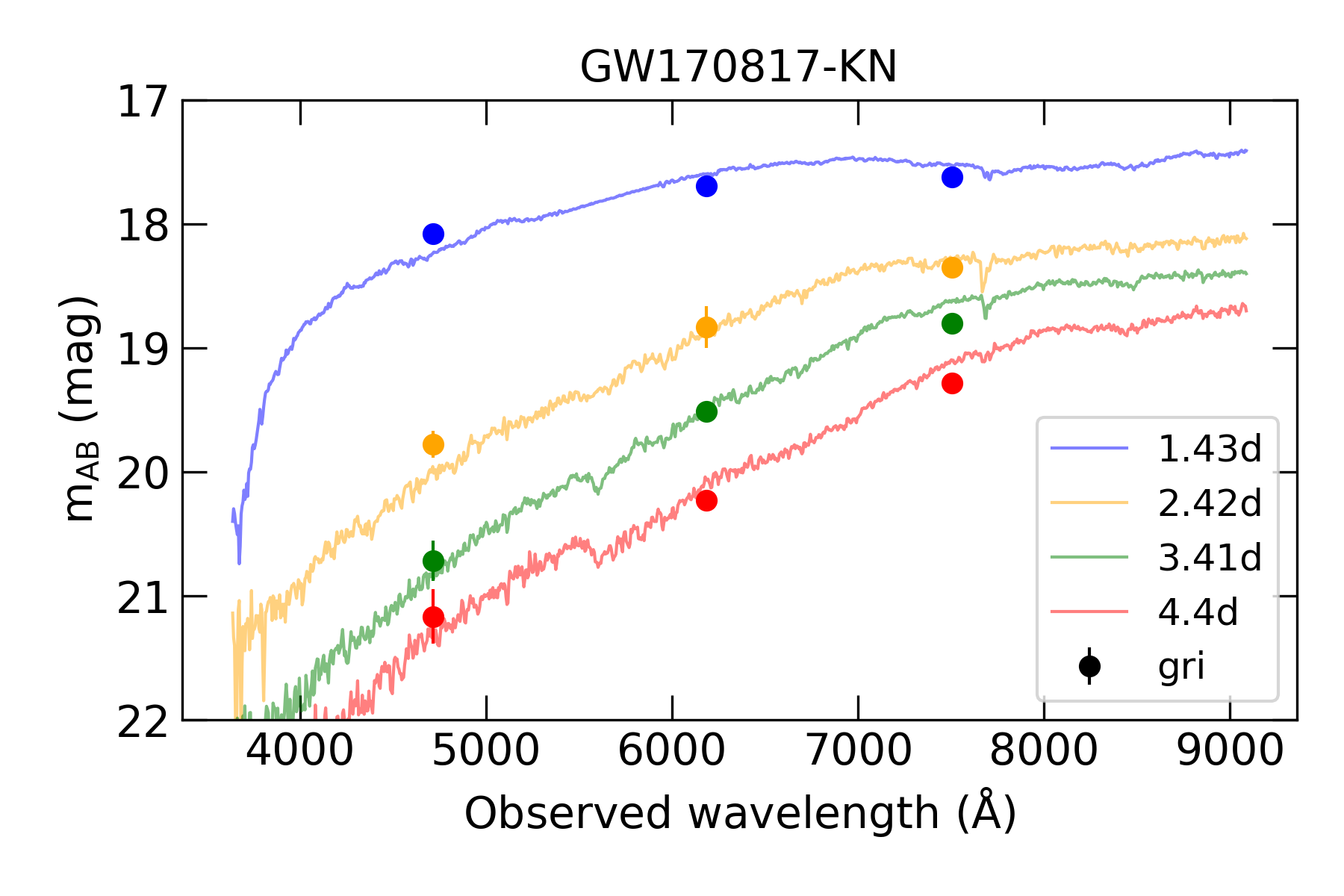}
    \caption{Time series of X-shooter AT 2017gfo spectra as if observed with MAAT with the OSIRIS R1000B and R1000R grisms for $0.5$ hr exposure per epoch. The observed broad-band SDSS $gri$ magnitudes are also shown with circular markers \citep{Smartt2017,Pian2017}. Note that AT 2017gfo was found in a nearby galaxy at $40$ Mpc and with an almost polar inclination angle direction $\sim$20$^{\circ}$ \citep{Hotokezaka2019}.}
    \label{fig:17gfoMAAT}
\end{figure}

To simulate GW170817 KN, we use the X-shooter spectra that were taken for AT 2017gfo \citep{Smartt2017,Pian2017} as our input templates and then compute how it would have been observed with MAAT { using our pipeline}. The spectra and the photometric data are publicly available\footnote{https://sid.erda.dk/wsgi-bin/ls.py?share\_id=bJ0ZS1CdFj}. 
The photometric data set for AT 2017gfo cover filters from the UV to the IR up to $\sim$10 days after the merger.  
In this analysis, we take the observed photometric data of AT 2017gfo in $gri$ bands on epochs  1.427, 2.417, 3.412, 4.402 days for a fair comparison with MAAT on wavelength range and timespan for our analysis.
Fig. \ref{fig:17gfoMAAT} shows the resulting time series spectra simulated for MAAT together with the overlap $gri$ observed photometry.
GW170817 happened at relatively close distance and the S/N per wavelength with MAAT would have been 140 at 7000$\AA$.

\begin{figure}
    \centering
    \includegraphics[width=.96\linewidth]{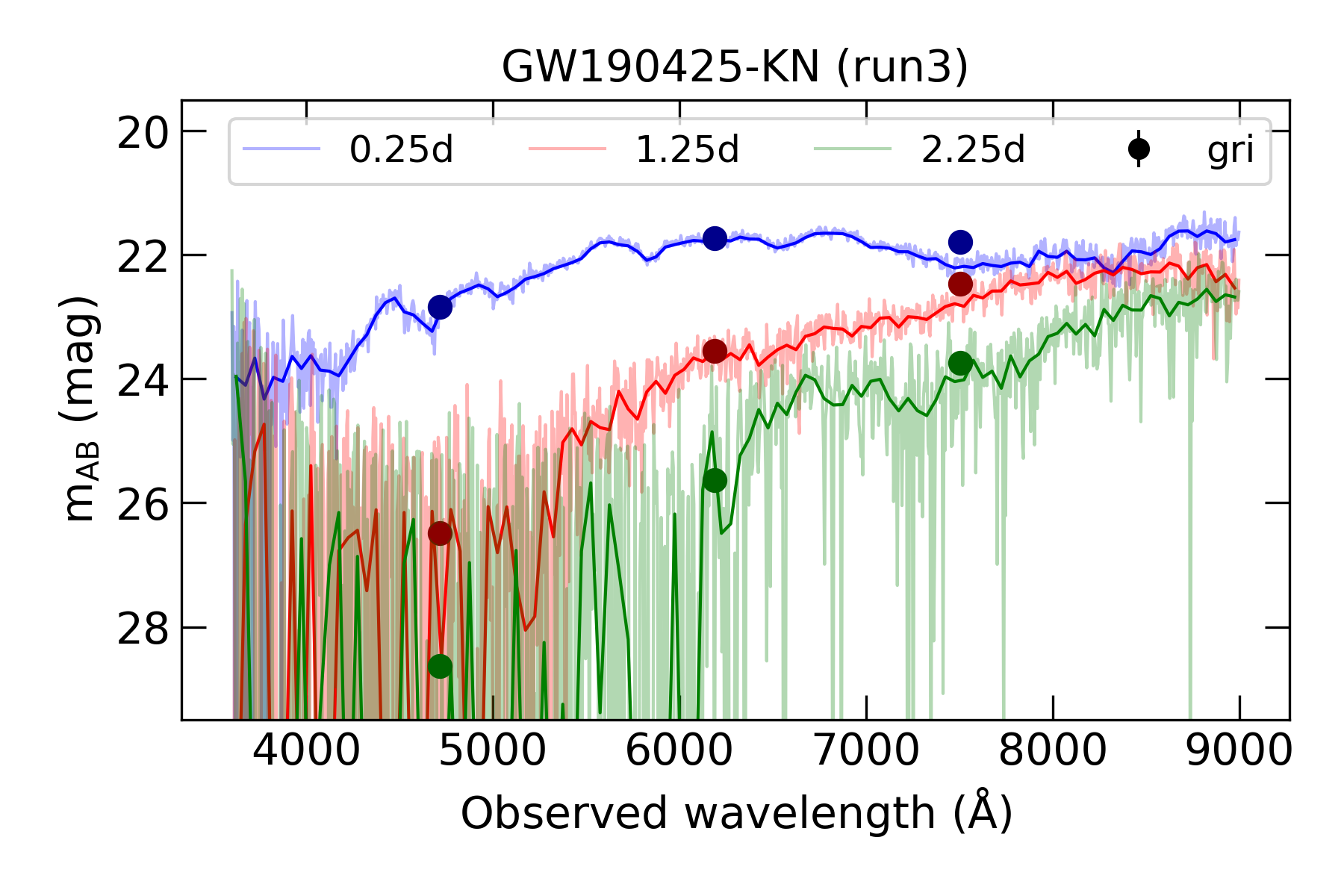}
    \caption{Time series spectra of a simulated KN (Run3 - DD2 EoS, see Table~\ref{tab1}) associated with GW190425 as if observed with MAAT with the OSIRIS R1000B and R1000R grisms for $0.5$ hr exposure per epoch. Note that the GW190425 event was located around $160$ Mpc and with an inclination angle $\sim40^{\circ}$. The broad-band SDSS $gri$ magnitudes are also shown, see text, with error corresponding to SNR$=$5, we notice that the observations have to be sensitive to 22 mag and deeper to get a detection for this kind of KN. The graph shows the MAAT spectra directly obtained from the ETC (shaded) and the rebinned spectra overplotted with solid line on the same color. Note that this is synthetic data generated for a KN at 159 Mpc and $\theta=40^{\circ}$.}
    \label{fig:190425MAAT}
\end{figure}

\begin{figure}
    \centering
    \includegraphics[width=.96\linewidth]{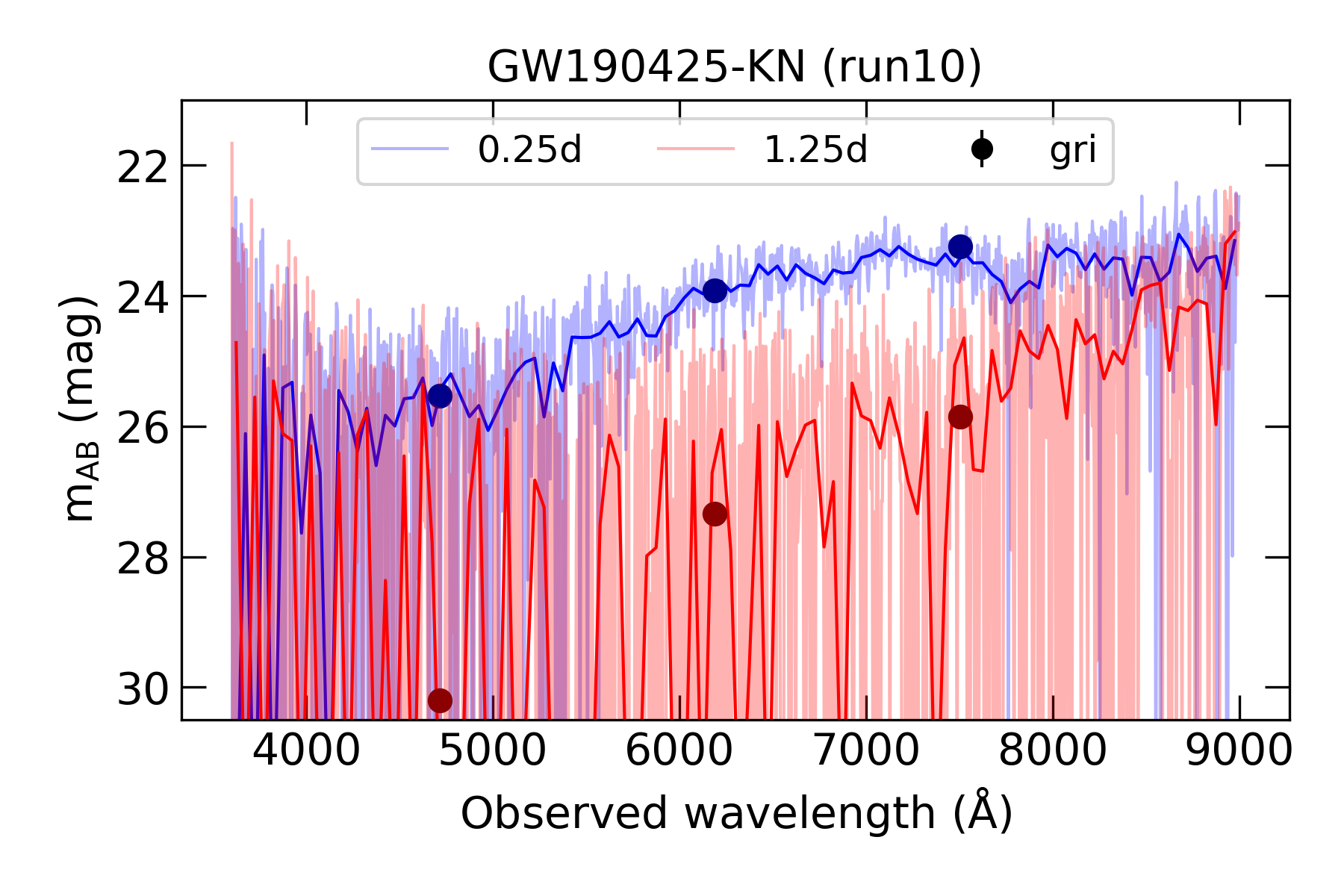}
    \caption{Same as Fig.~\ref{fig:190425MAAT}, but adopting a KN model with similar merger parameters but different EoS, i.e. run10 - LS220 EoS. This produces fainter KN spectra, whom significantly impacts the SNR of the MAAT spectra, and hence the constraining power on $H_0$. Nevertheless, thanks to the spectro-photometry capability and spectral resolution of MAAT, the SNR is enhanced by rebinning the spectra. The solid line shows the rebinned spectra in 50 $\AA$ and the marker the convolved photometry in $gri$ assuming $\rm SNR=5$. }
    \label{fig:190425MAAT_r10}
\end{figure}

\subsubsection{GW190425 KN light curves}

No KN was identified for GW190425. From the GW data, the luminosity distance was inferred to $159^{+69}_{-72}$ Mpc, and the inclination angle was poorly constrained with a median value of $\sim40^{\circ}$ \citep{abbott190425}. For our analysis, we simulate synthetic KN data assuming $D_L=159$ Mpc and $\theta=40^{\circ}$. We consider two of the models presented in this study: run3 and run10 (see Table \ref{tab1}) which chirp mass ($M_{chirp}$) and mass ratio ($q$) are consistent with GW190425. Run3 assumes DD2 EoS and run10 LS220, and the resulting KN varies in ejecta parameters such as the mass of the ejecta components, the electron fraction ($Y_e$), and the expansion velocity of the ejecta ($v_{ej}$).
{ The spectra of GW190425 have been simulated using the run3 and run10 described previously (see Table \ref{tab1}). These two runs are tailored to a BNS merger with a chirp mass ($M_{chirp}$) and mass ratio ($q$) consistent with GW190425.} 
Run3 produces a KN with $M_{\rm wind}=5.88\times 10^{-3} M_{\odot}$ and $M_{\rm dyn}=1.2\times 10^{-3} M_{\odot}$ while run10 generates a KN with $M_{\rm wind}=2.10\times 10^{-4} M_{\odot}$ and $M_{\rm dyn}=3.0\times 10^{-4} M_{\odot}$.
 For the same BNS, $q, M_{chirp}$ parameters with $M_{1}=1.6M_\odot$ the more compact i.e. smaller radius realization, see Figure~\ref{Fig:q_mc_m1}, run 10 LS220 EoS produces fainter light curves and then larger apparent magnitudes. EoS differences thus translate into significant changes in the luminosity and evolution of the light curves. For the KNe simulated for GW190425 we find that DD2 produces brighter and longer-lived KN in the wavelength range of MAAT.  

Figure~\ref{fig:190425MAAT} shows the resulting time series spectra for run3, while Figure~\ref{fig:190425MAAT_r10} the same for run10. For both cases, the spectral series are much noisier than the one for AT 2017gfo. For run3, the SNR estimated at 7000$\AA$ goes from $\sim8$ lowering to $\sim2$ and $\sim1$ within 2 days, but at later epochs we report an increase of the SNR at red wavelengths, as it is expected from the KN evolution. Run10's spectral series shows even fainter fluxes (the SNR peaks at around 1 and drastically decreases) with magnitudes shallower than 22-23 mag, which would be hard to detect with MAAT, and even with current optical surveys. If MAAT followed up on such KN, we would be able to gather useful information only by rebinning the spectrum (Fig \ref{fig:190425MAAT_r10}), which improved the SNR by a factor of 5. The broadband photometry, in this case, would need to be sensitive to 23-24 apparent magnitudes. The incoming Rubin Observatory Legacy Survey of Space and Time (LSST, \citealp{LSST}) will be able to detect future KNe at these fluxes.

As there is no EM counterpart associated with this event, therefore, we choose to simulate synthetic $gri$ photometry with standard errors for optical depth $\sim22$ mag. We aim to compare MAAT observations with what would have likely been observed by optical surveys if our assumed KN were detected. 
Run3 and run10 KNe at 159 Mpc would be hard to observe with optical surveys reaching $\sim22$ mag limiting magnitudes as the sky background would largely dominate (S/N$>5$).
We plan to explore synthetic photometry that resembles more dedicated follow-up observations in the future, therefore, reaching greater sensitivity at shallower magnitudes. 



\subsection{Measurement of the Hubble constant}
As noted above, there is a $5\sigma$ tension between the inferred value of $H_0$ from the cosmic microwave background \citep{Planck2020} and the local Cepheid distance ladder \citep{Riess2021}. This tension can be a sign of new physics beyond the standard cosmological model or unknown systematics. BNS mergers with an identified EM counterpart \citep[e.g.][]{distance} are an excellent probe to understand the origin of this tension since they measure $H_0$ completely independently of the distance ladder, or to any assumptions on the cosmological model.


The use of GW as "standard sirens" to estimate $H_0$ represents a powerful distance indicator in cosmology  \citep{Holz2005}. The luminosity distance is calculated directly from the GW signal, despite it is degenerate with the inclination angle ($\theta$). However, it has been proposed to use the electromagnetic data to constrain the inclination angle and break the degeneracy \citep{Nissanke2013arXiv1307.2638N}. 

{ The standard siren method can also be used in the case the exact identification of the galaxy hosting a BNS event is missing. This approach, originally proposed by \citet{Schutz1986}, takes into account all the bright host galaxies located in the GW localization region as potential host galaxy of the BNS merger event, and then infer the H$_0$ value as the average value obtained from the ensemble analysis. A similar approach has been applied for the case of GW170817, where a value of $H_0 = 77^{+37}_{-18}$ km/s/Mpc has been inferred \citep{Fishbach2019}.} However, in the case of detecting an EM counterpart, it is possible to unambiguously identify a host galaxy and infer a redshift to the source. Moreover, the EM counterpart also contains essential information regarding the inclination of the binary system. This possibility has been explored with the KN and the GRB associated with GW170817. Here, we infer those constraints by simulating how such a KN would be observed with MAAT IFU spectroscopy on the GTC telescope. Along with simulations for AT 2017gfo, we also project the improvements for the well-observed GW event GW190425. It will depend on factors like the S/N of the observed spectrum and the shape of possible features related to the geometry and composition of the KN ejecta.
We join GW and KN data as presented in \cite{Dhawan2020}, by taking the joint posterior probability distribution in $H_0$ and $\cos(\theta)$ from the GW data and applying a prior on the inclination angle from the KN. 

{ The methodology followed in our analysis is explained in what follows. We use the synthetic data from MAAT and the photometry for each BNS (GW170817 and GW190425) and perform parameter inference using simulations with the radiative transfer algorithm \texttt{POSSIS} \citep{Bulla2019}. The code predicts 3D models for a set of ejecta parameters and geometries including the inclination relative to the observer. The predicted viewing angle dependence allows us to infer the $\cos(\theta)$ parameter from the KN data independently of the GW. A set of models from \texttt{POSSIS} is considered, including the grid presented in \cite{Dietrich2020} and the 23 models presented in this study (see Section \ref{sec:modelsandEOS} and Table~\ref{tab1}). To infer the KN parameters, we perform $\chi^{2}$ fits and extract the $\chi^{2}$ probability distribution of the viewing angle parameter marginalizing over the ejecta parameters, which include the masses of the ejecta components and the half opening angle of the lanthanide-rich region. The obtained viewing angle distribution is used as a prior for $\cos(\theta)$ to reweight the GW joint posterior from the GW analysis. The extra and independent constraint from the KN data on $\cos(\theta)$ breaks the degeneracy with the luminosity distance and results in an improvement in the $H_0$ uncertainty with respect to GW data alone. This method is used in this section to analyze the contribution of viewing angle inference with KN observed with MAAT to the $H_0$ constraints.  
 }

\begin{figure*}[ht!]
\begin{center}
  \includegraphics[width=0.90\linewidth]{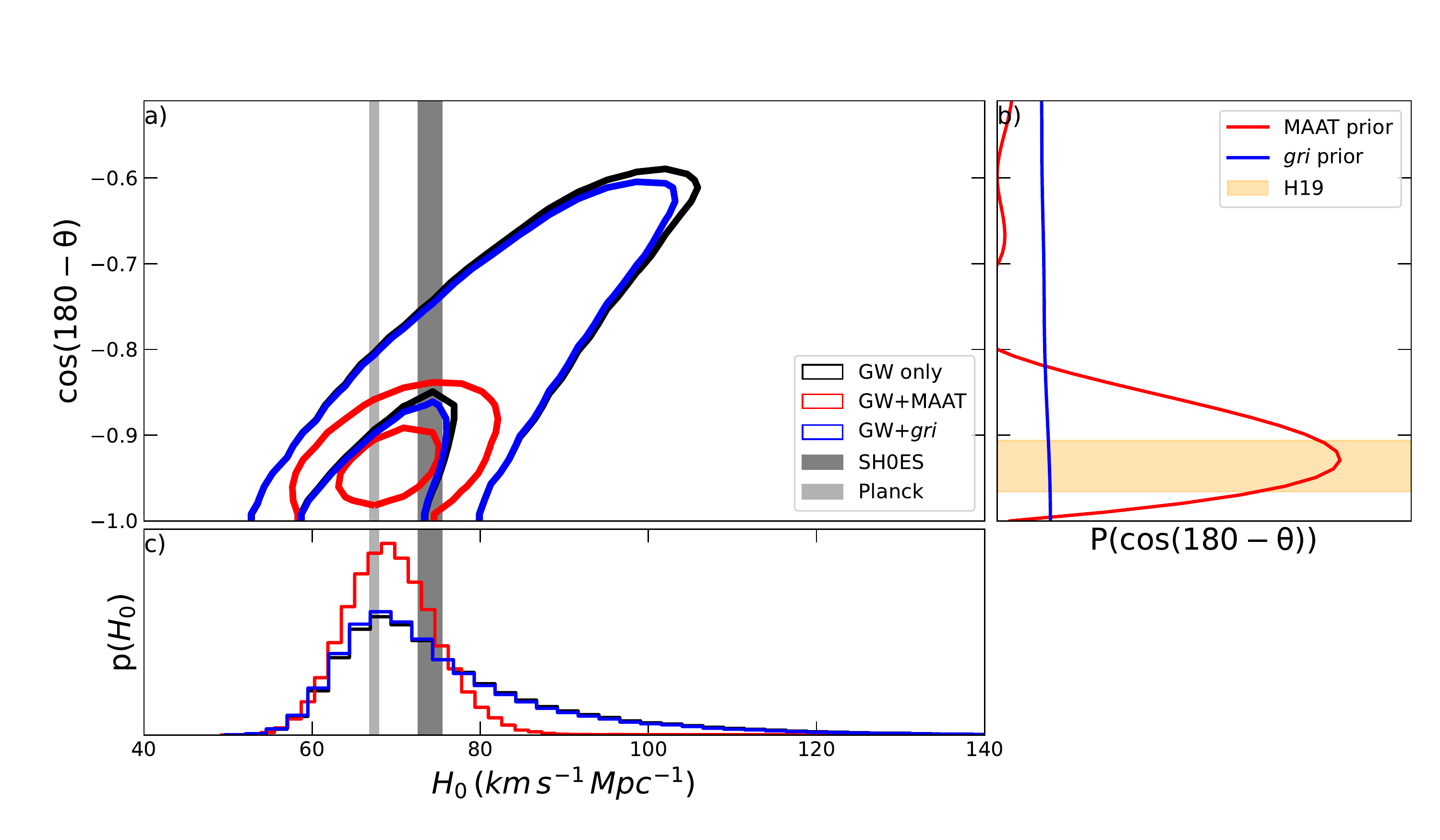}
   \caption{Hubble constant determination with prior distribution from GW170817. a) shows the 1 sigma and 2 sigma contours on the parameter space on inclination angle and $H_0$ with no KN constraint (black) and with KN constraint from MAAT data (red) and broadband $gri$ photometry (blue). b) shows the EM priors applied to the GW posterior extracted from MAAT data (red) and $gri$ AT 2017gfo observed photometry (blue). c) The marginalization on $H_0$ for the GW posterior (black) and the reweighted posterior with MAAT prior (red) and $gri$ prior (blue) applied. This event occurred at a luminosity distance of 40 Mpc, corresponding to a redshift of around 0.009.}
  \label{figH0_17gfo}
\end{center}
\end{figure*}
\subsubsection{$H_{0}$ estimate from GW170817}

For the case of GW170817, we take the posterior distribution presented in \cite{abbott2017a}. The $H_0$ at joint probability distribution for this event was calculated by combining the luminosity distance with the redshift measured for the host galaxy NGC 4993 of AT 2017gfo. 

The prior inclination angles are calculated following a $\chi^2$ distribution that indicates the goodness of fitting our {simulated AT2017gfo MAAT spectrum, shown in Fig.~\ref{fig:17gfoMAAT}, with the BNS  model grid from the radiative transfer code \texttt{POSSIS} presented in \cite{Dietrich2020}.} 
 The resulting inclination angle distribution is shown in the red curve of panel b) in Fig.~\ref{figH0_17gfo}. The $\theta$ constraint is significant to the order of 10 percent, and the $H_0$ uncertainties improve by a factor of $\sim$40$\%$ with respect to GW data only (see Table \ref{h0_results}). 
As a comparison, it is also shown in Fig.~\ref{figH0_17gfo} a yellow band corresponding to the constraint obtained from the superluminal motion of the associated short GRB jet in \cite{Hotokezaka2019}.

The same analysis has been performed for the photometric data of AT 2017gfo {introduced in Section~\ref{subsec:17gfoLC}}. \cite{Dhawan2020} obtained a 24$\%$ improvement with the full UV-optical-IR dataset. We analyze only data on the same epoch and wavelength range as the MAAT spectra (3600-9000 $\AA$) with  $gri$ filters and 4 epochs: 1.427, 2.417, 3.412, and 4.402 days. The {final $H_0$} constraint from {GW plus} the described photometry improves only by a factor of 5\% {which compares to the $\sim40\%$ from GW+MAAT spectra}, probing the benefit of MAAT data in KN parameter inference for the wavelength range (see Table \ref{h0_results}). 

\begin{table}[ht!]
\centering                                      
\resizebox{0.5\textwidth}{!}{\begin{tabular}{c c c c c c }         
\hline\hline                        
 BNS       &  KN               &    &    $\sigma(H_0)/\sigma(H_{0,\rm GW-only})$ \\    
\hline                                   
 GW170817  &  ---   & GW only           & 1.0 &   \\    
 GW170817  &  AT 2017gfo data       & GW+$gri$ prior    & 0.95 &   \\    
 GW170817  &  AT 2017gfo template       & GW+MAAT prior     & 0.54 &   \\      
 GW190425  & ---     & GW only           & 1.0 &   \\    
 GW190425  & Run 3 (DD2)     & GW+$gri$ prior    & 0.85 &   \\    
 GW190425  & Run 3 (DD2)     & GW+MAAT prior     & 0.73 &   \\
 GW190425  & Run 3 (DD2)     & GW+MAAT (50$\AA$) & 0.60 &   \\
 GW190425  & ---                 & GW only           & 1.0 &   \\    
 GW190425  & Run 10 (LS220)      & GW+MAAT prior     & 1.0 &   \\
 GW190425  & Run 10 (LS220)      & GW+MAAT (50$\AA$) & 0.84 &   \\
%
\hline                                           
\end{tabular}}
\caption{$\sigma(H_0)$ ratio results. Note that we are not paying attention to the $H_0$ value in this analysis, but we focus mainly on the resulting uncertainty improvement to GW-only. 
} 
\label{h0_results}      
\end{table}

\begin{figure*}[ht!]
\begin{center}
  \includegraphics[width=0.96\linewidth]{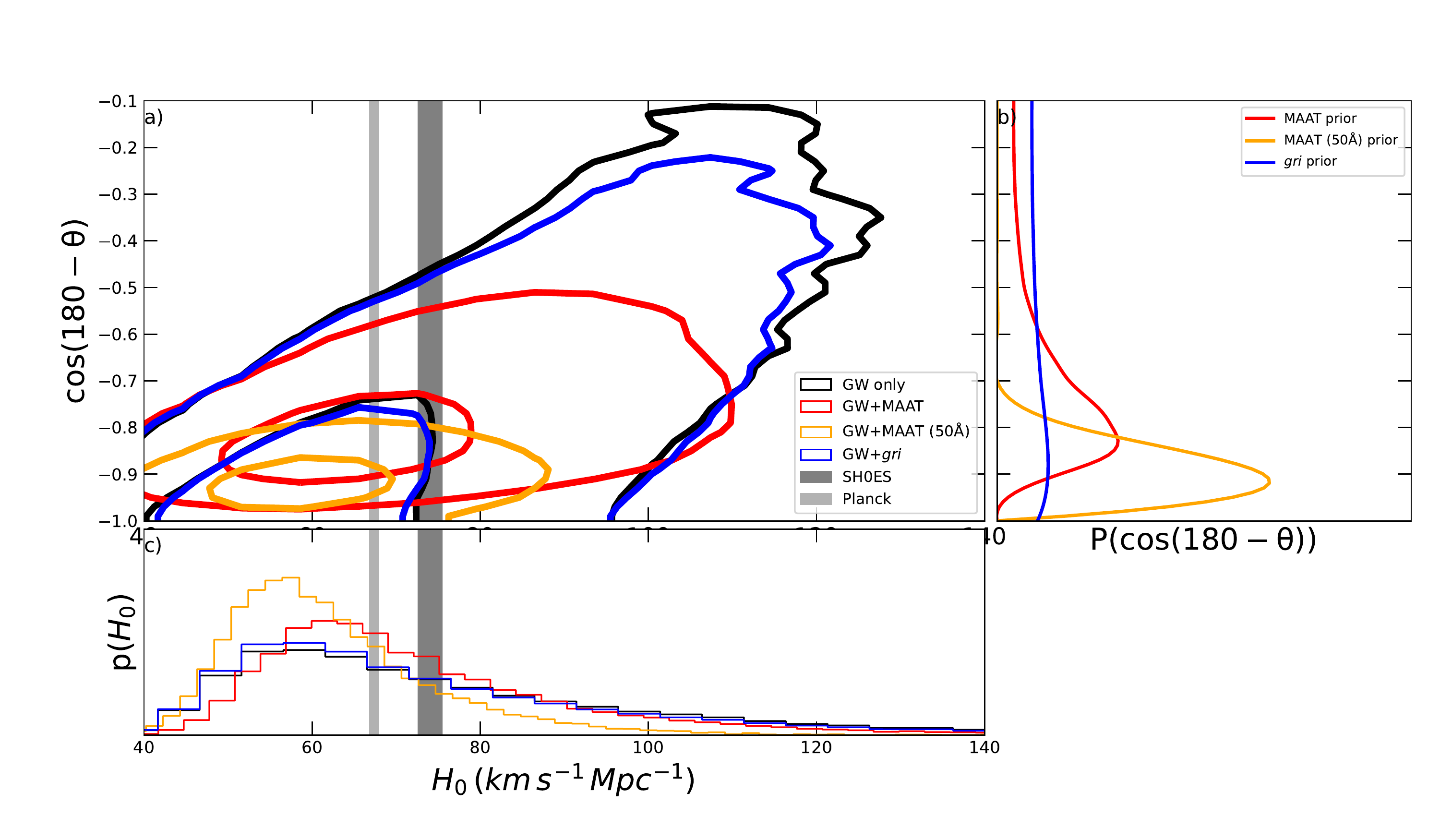}
   \caption{Hubble constant determination with prior distribution from GW190425. This figure follows the same panel structure as Fig.~\ref{figH0_17gfo}. The orange contour adds the constraint from a rebinned spectrum of 50$\AA$. The synthetic KN used to compute this plot assumes merger parameters consistent with GW190425 and EoS DD2 (run 3, see Table \ref{tab1}) a luminosity distance of 159 Mpc ($z\sim0.035$) and a inclination angle of $\sim40^{\circ}$. Note that $gri$ prior is computed for a survey with depth $\sim22$ mag while the MAAT 50$\AA$ prior comes from a rebinning of the observed MAAT spectra, which improves SNR. }%
  \label{figH0_run3}
\end{center}
\end{figure*}

\subsubsection{$H_{0}$ estimate from GW190425}

For GW190425 we generate the $H_0$ - $cos(\theta)$ joint posterior from the luminosity distance and assuming a redshift $z=0.035$  \citep{abbott190425}. Note that the central value of the $H_0$ distribution is not physically meaningful as it purely depends on the chosen redshift input, while the shape indicates the actual uncertainties. We lack redshift information for GW190425. We cannot correct for recession velocity with no EM counterpart. Therefore, we cannot measure the real $H_0$ value using this method. Nevertheless, we assume a redshift of 0.035 
for the synthetic GW190425 - KN and apply the same to the GW luminosity distance data from which we estimate $H_0 \approx ( c \cdot z )D_L $, with $c$ the speed of light, $z$ the redshift and $D_L$ the luminosity distance. This analysis aims to quantify the constraining power of MAAT, for what the relevant results are the uncertainties in the $H_0$ as the improved constraint with KN compared to GW only. We constrain the $\theta$ parameters that we present and described above in the paper. Regarding the simulated photometry of GW190425, we assume $gri$ SDSS filters and a 5 sigma detection at 22 mag. 
The SNR for such photometry goes below 5 (no detection), which significantly lowers the constraining power, as we can see in the blue curves in Figure~\ref{figH0_run3}.

The $H_0$ analysis on the synthetic KN data highlights the potential of MAAT on improving the constraints even for more distant KNe. If we assume that the models considered are representative of KNe in the Universe, even one single spectrum with the resolution of the R1000 grisms can significantly constrain the $\theta$ parameter and improve $H_0$ by 28\%, or even further when using  "cleaner" data like the 50 $\AA$ binned spectra shown in orange in Fig.~\ref{figH0_run3}, providing a 40\% improvement (see Table \ref{h0_results}) with respect to GW data only. Our results on the $H_0$ for GW190425 depend on the assumed KN model and assumed redshift, therefore, we do not intend to provide a realistic $H_0$ value for GW190425, but only to show the improved uncertainties and the dependence on the model assumptions, including the chosen EoS.  

We did the exercise with the simulated KN assuming EoS LS220 (run10), which shows spectral noise after the first epoch as the background flux dominates after day 1. Suppose observing 0.5 hr with gratings R1000 (Fig. \ref{fig:190425MAAT_r10}, we are not able to constrain the KN parameters with MAAT spectra due to the poor S/N, nevertheless with a rebinning of 50$\AA$, we obtain a factor of $\sim$5 better SNR and the inclination angle prior provides a 15\% improvement in the $H_0$ uncertainties (see Table~\ref{h0_results}).  

We note that for KNe at $\sim$150 Mpc, it is critical to obtain early data to constrain our model parameters. In the case of run3, most of the constraint comes from the first epoch assumed to be taken within the first day after the merger, indicating the need for fast response efficient communication of KN candidates. The constraining power of our data depends on the S/N, the consequence of the observing strategy, the intrinsic brightness of the KN, the luminosity distance, and the orientation, among other factors. The ability to constraint model parameters is also highly dependent on the model grid, as our models need to be able to reproduce the main features of the KN to provide sensitive results. This study assumes that our models represent the produced KN emission. 

We aim to understand if the EoS induces systematics in the measurement of the Hubble constant. This study only compares run3 and run10 (described above) to simulate a KN associated with GW190425 with two different EoS: DD2 and LS220. We did the exercise to simulate MAAT data for both runs but requiring the same S/N of 5 at 6000$\AA$, finding that the pipeline returns similar prior to the inclination angle in both cases, which indicates that the constraining power is mainly dependent on the SNR for these types of KN. 

We plan a follow-up study to investigate the effect of the EoS in H0 precision for a population of KNe.   
In practice, it will not be possible to obtain the same S/N at all luminosity distances, as we need to consider the exposure time needed. We compare the inclination angle priors and $H_0$ estimation for run3 and run10 with the same exposure time of 1 hr for the KNe located at 100 Mpc (still consistent with GW190425 within 1 sigma). Run3 SNR is 4 times larger than run10, which is around 5 at 6000$\AA$. 








\section{Discussion and conclusions}
\label{conclusions}
KN multimessenger emission produced in the BNS merger can shed light on fundamental unresolved questions, such as the behaviour of matter at high densities inside NSs or the precise determination of the Hubble constant, $H_0$. In this work, we show that obtaining spectro-photometric light-curves of future KNe will allow us to constrain $H_0$ more precisely than has been done previously from GW-only or using broad-band light curves. This paves the way for a yet more precise complementary measurement of the Hubble constant. We also show how KN peak magnitudes are correlated to NS properties that closely depend on the EoS. In particular, stellar radii and compactness (M/R ratio) are key to understand the amount of matter ejected, and thus the peak brightness. In our study we find an indication that there are limiting values of compactness to obtain measurable light curves.

 Using the 3D radiative transfer code \texttt{POSSIS} \citep{Bulla2019}, we predict the KN signal for a set of input models from \cite{nedora} and \cite{Radice}. We analyze the dependence of different physical parameters governing the KN emission, i.e. ejecta properties specific to the particular BNS configuration based on mass ratio of NS masses, $q$, and chirp mass $M_{\rm chirp}$ as well as inclination angle and cosmological distance (provided by GW measurements). 
 
 We detail how MAAT on the GTC, a new optical integral field unit that employs the image slicing technique to perform absolute flux calibrated spectroscopy in a field of view of 8.5$^{\prime \prime}$ $\times$ 12.0$^{\prime\prime}$, is well suited to shed light on the characteristics of the optical emission of KNe via follow-up of BNS events.

We find that the signal modulation as a result of the inclination angle must be treated in a systematic fashion as it can result in up to one order of magnitude difference at peak luminosity. Fainter events experience a more rapid evolution of the bolometric luminosity seven days after the merger. The brighter the KN the later it peaks, with a consistent trend of peak times within $\sim 10$ days. It is for this reason that setting an early observation strategy is key for capturing the essential physics governing the transient event. As we have explained, the underlying EoS describing the interior of individual NSs in the binary with given ($q,M_{\rm chirp}$) determines the amount and properties of ejecta. We consider a reduced but representative set of EoS, namely DD2, SFHo, and LS220. We find that the light curve for times $t<t_{peak}$ can yield sizable differences, up to one order of magnitude, thus showing the importance of a rapid follow-up for these transients. 

We further study the dependence of the luminosity peak with average ejecta mass, velocity, and electron fraction as obtained from NR samples. We find a strong dependence between dynamical plus disk mass ($M_{dd}$) and chirp mass for $q=1$ events; and weaker for $q>1$ events. 
Even if the disk mass does not play an effective role in the KN EM emission (only indirectly as a source of the wind mass), it is the bulk of matter originating from the layers less gravitationally bound to the NSs and thus susceptible of being tidally ejected during the violent merger event. 
We find that the binary tidal polarizability (yet another parameter available from GW observables; presently constrained from GW170817 to be $\tilde{\Lambda}\lesssim 800$) can effectively limit the amount of mass ejected (for a given EoS) and constrain the peak luminosity. From our initial analysis, an  upper bound log $L_{peak}[\rm erg/s]\sim 41.3$ seems to indicate that there is yet another complementary constraint on the combined NS mass – radius diagram. From the set of EoS analyzed we find that some EoS are not capable of producing such bright KNe as measured for AT 2017gfo.

The resulting KN from the NS coalescence is  mostly dictated by how nuclear matter behaves and the degree of compactness attainable for a NS. We plan to perform a more quantitative study on this matter by simulating KN for a given GW event at a reasonable luminosity distance such that we can compare the analysis and understand the implication that the EoS has on the optimal observing strategy, and ultimately on the $H_0$ estimation. 

We predict the improvement of the $H_0$ estimate with MAAT data for a $\sim22$ limiting magnitude survey. We conclude that fitting the light-curves over the whole spectral range (3600–10000 $\AA$) provides a better constraint than that from using GW-only and broadband ($gri$) photometric data. The resulting improvement on $H_0$ would have been around 40\% if MAAT would have observed AT 2017gfo.
This compares to other analysis found in the  literature like that of \citep{Dhawan2020} who obtains a 25\% improvement from the full UV-Optical-IR dataset during 10 days after the merger. 
GW190425, the possible KN at $\sim$159 Mpc, is expected to be much fainter than AT 2017gfo because of the larger distance and less favorable inclination angle of around 40$^{\circ}$ \citep{abbott190425}. 

The resulting constraint in $H_0$ largely depends on the EoS assumed, as it impacts mostly on the brightness of the KN emission. We investigated the effect of using two different EoS (DD2 and LS220) for a merger like GW190425 with $M_{\rm chirp} = 1.39 \, M_{\odot}$, $M_1 = 1.6 \, M_{\odot}$, and $q=1$. At 159 Mpc, the run10 KN observed with MAAT is not constraining unless rebinning (increase SNR) is done, what leads to an improvement of $\sim15\%$ mainly from the first epoch obtained within the first day (see Table \ref{tab1}). Run3 KN instead shows a better improvement on the $H_0$ precision of $\sim25\%$ to $\sim40\%$ with rebinning (see Fig. \ref{fig:190425MAAT} and Table \ref{tab1}). 

\begin{figure*}[ht!]
\begin{center}
  \includegraphics[width=0.90\linewidth]{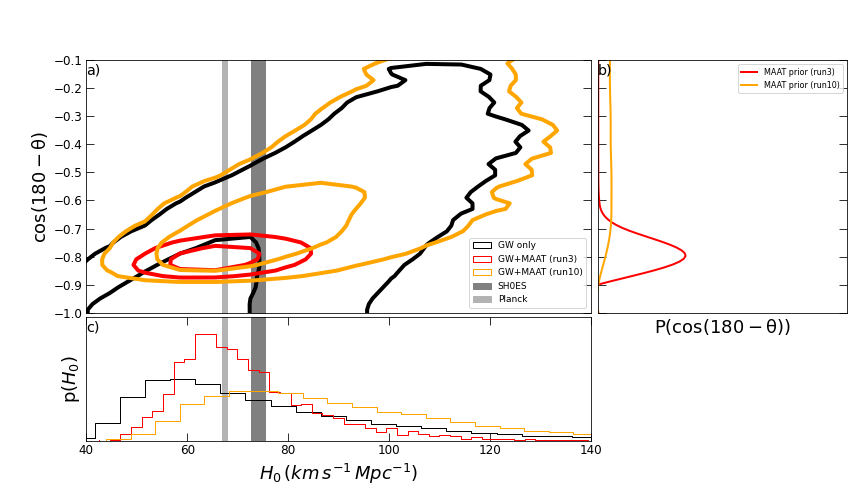}
   \caption{Hubble constant determination with prior distribution from GW190425 comparing results for run3 EoS DD2 (red) and run10 EoS LS220 (orange), assuming GW190425-KN at 100 Mpc redshift 0.022 and 1 hr exposure time. This figure follows the same panel structure than Figures~\ref{figH0_17gfo} and ~\ref{figH0_run3}. 
  }
  \label{figH0run310}
\end{center}
\end{figure*}

We also investigated for a KN at 100 Mpc arising from the same merger event but assuming two different EoS (again DD2 run3 and LS220 run10) with the same S/N (S/N$=5$ at 6000$\AA$) or same exposure time (1 hr). The results show that for a consistent S/N, if the two models considered are representative, we are able to constrain $H_0$ similarly, while for the same exposure time the constraining power shows a clear dependence on EoS (see Fig. \ref{figH0run310}) due to the much fainter magnitude and poorer S/N of LS220 run10. We plan a detailed follow up study to understand possible systematics and bias on $H_0$ introduced from EoS uncertainties and other properties. The possibility to get early KN spectra with a high S/N with MAAT will also facilitate the identification and the study of spectral features embedded within the KN ejecta, paving the way for abundance studies with spectral synthesis codes, and allowing us to neatly determine the kinematics of KN components.

\begin{acknowledgements}

We thank M. Branchesi, M. Tanaka and S. Smartt for useful comments, and S. Rosswog and O. Korobkin for sharing heating rate libraries and corresponding interpolation formulae.

MAPG and CA acknowledge financial support from Junta de Castilla y Le\'on through the grant SA096P20, Agencia Estatal de Investigaci\'on through the grant PID2019-107778GB-100. DB acknowledges support from the program Ayudas para Financiar la Contrataci\'on Predoctoral de Personal Investigador (Orden EDU/1508/2020) funded by Consejer\'ia de Educaci\'on de la Junta de Castilla y Le\'on and European Social Fund. LI, AA, CRA, CG and JH were supported by grants from VILLUM FONDEN (project number 16599, 25501 and 36225). The Oskar Klein Centre authors acknowledge support from the G.R.E.A.T research environment, funded by {\em Vetenskapsr\aa det}, the Swedish Research Council, project number 2016-06012. MB acknowledges support from the Swedish Research Council (Reg. no. 2020-03330). The IAA-CSIC authors acknowledge financial support from the State Agency for Research of the Spanish MCIU through the ``Center of Excellence Severo Ochoa'' award to the Instituto de Astrof\'isica de Andaluc\'ia (SEV-2017-0709). EP and CDT thank support from the Agencia Estatal de Investigaci\'on through the grant AYA2016-77846-P. FP and CDT thank the support of the Spanish Ministry of Science and Innovation funding grant PGC2018-101931-B-I00.CDT thanks the Instituto de Astrof\'isica de Canarias for the use of its facilities while carrying out this work. DJ acknowledges support from the Erasmus+ programme of the European Union under grant number 2020-1-CZ01-KA203-078200. SD acknowledges support from the Marie Curie Individual Fellowship under grant ID 890695 and a junior research fellowship at Lucy Cavendish Fellowship. AA’s research is funded by Villum Experiment Grant \textit{Cosmic Beacons} project number 36225.

The MAAT instrument project consists of a partnership of member institutes in Spain (Instituto de Astrof\'isica de Andaluc\'ia CSIC, Instituto de Astrof\'isica de Canarias), Denmark (DARK, Niels Bohr Institute, University of Copenhagen), and Sweden (The Oskar Klein Centre for Cosmoparticle Physics, Stockholm University).

\end{acknowledgements}

\bibliographystyle{aa}
%

\bibliography{biblio}

\end{document}